\documentclass[journal]{elsarticle}
\usepackage{graphicx,bm,epsf,float,amssymb}
\usepackage[font={scriptsize}]{subcaption} 

\usepackage{lineno}
\usepackage{multirow}
\usepackage{graphicx}

\begin{document}
\begin{frontmatter}
\title{Characterization of a silicon photo-multiplier array with summing board as a photo-multiplier tube replacement in organic scintillator assemblies}

\author[sandia-ca]{M. Sweany\corref{cor1}}
\author[sandia-ca]{P. Marleau}
\author[awe]{C. Allwork}
\author[sandia-nm]{G. Kallenbach}
\author[sandia-nm]{S. Hammon}

\address[sandia-ca]{Sandia National Laboratory, Livermore, CA 94550, USA}
\address[sandia-nm]{Sandia National Laboratory, Albuquerque, NM 87185, USA}
\address[awe]{Atomic Weapons Establishment, Reading, UK}
\cortext[cor1]{Corresponding author: msweany@sandia.gov}

\begin{abstract}
We report on the energy, timing, and pulse-shape discrimination performance of cylindrical 5 cm diameter x 5 cm thick and 7 cm diameter x 7 cm thick  {\it trans}-stilbene crystals read out with 
the passively summed output of three different commercial silicon photo-multiplier arrays. Our results indicate that using the summed output of 
an 8x8 array of SiPMs provides performance competitive with photo-multiplier tubes for many neutron imaging and correlated particle measurements: for the 5x5 cm crystal read out with 
SensL's ArrayJ-60035\_64P-PCB, which had the best overall properties, we measure the energy resolution 
 as 13.6$\pm$1.8\% at 341 keVee, the timing resolution in the 100--400 keVee range as 277$\pm$34 ps, and the pulse-shape discrimination figure-of-merit 
as 2.21$\pm$0.03 in the 230--260 keVee energy range. These results enable many scintillator-based instruments to enjoy the size, robustness, and power benefits of silicon 
photo-multiplier arrays as replacement for the photo-multiplier tubes that are predominantly used today.
\end{abstract}


\end{frontmatter}

\section{Introduction}
One of the most ubiquitous radiation detection systems consists of a photo-multiplier tube (PMT) coupled to a scintillating material, such as organic scintillator or one of many scintillating inorganic 
crystals. While widely used for a variety of applications, the size, power requirements, and robustness of such a system is fundamentally limited by that of the PMTs. One alternative is micro-channel 
plate PMTs (MCP-PMTs), which are smaller and therefore more robust to magnetic fields. However, they still require high voltage operation (typically above 1kV), require a vacuum seal, and are much 
more expensive than traditional PMTs. Recently, silicon photo-multipliers (SiPMs) have demonstrated the necessary gain and dark current levels for single photon detection, potentially enabling them 
to replace PMTs in many applications. The timing response has been demonstrated to be sufficient for particle identification via pulse-shape discrimination when coupled to organic scintillators 
\cite{ruch,grod,yu}.  The price of commercially available devices is also competitive with PMTs of similar photo-cathode coverage. SiPMs have other features that improve the robustness and useability 
compared to traditional PMTs: they are compact, operate on low voltage (typically under 100V), and, like MCP-PMTs, they are insensitive to magnetic fields. The photo-detection efficiency (PDE) is 
typically higher than PMTs, with recent devices having a maximum $\sim$50\% PDE. The PDE is also much less dependent on the incident position or angle of the photon compared to PMTs: typical 
PMT responses at 60$^\circ$ incidence drops to approximately 60\% of the value at zero incidence \cite{hamamatsu}, while recent measurements on SiPMs report a constant 
response as a function of incidence angle out to approximately 65$^\circ$ \cite{romeo}. 

In this work, we report on the development of a SiPM-based readout for a 5 cm diameter x 5 cm thick {\it trans}-stilbene crystal from Inrad Optics, and compare the energy, timing, and pulse-shape 
discrimination from the summed output of three  different SiPM arrays. These metrics were chosen for applicability to neutron kinematic imaging  \cite{pirard, riviere, miner} and fast neutron-gamma 
multiplicity analysis \cite{toffee} applications. We use the 8x8 array of 6x6 mm$^2$ C-series SiPMs from SensL (ArrayC-60035\_64P-PCB), the 8x8 array of 6x6 mm$^2$ J-series SiPMs also from 
SensL (ArrayJ-60035\_64P-PCB), and a 2x2 array of Hamamatsu's S13361-6050 4x4 arrays, closely packed on an adaptor board to create an 8x8 array. Each array has 
approximately the same cross section, however there are small differences due to the separation between pixels: the SensL C-series array is 5.7x5.7 cm$^{2}$, and the J-series and Hamamatsu 
arrays are 5x5 cm$^{2}$. Each array has at least four pixels that do not overlap with the 5 cm diameter crystal due to its rectangular cross section. 

\section{Passive Summing Board}
\begin{figure}[!htbp]
		\centering
		\begin{subfigure}[h]{\columnwidth} 
			\includegraphics[width=\columnwidth]{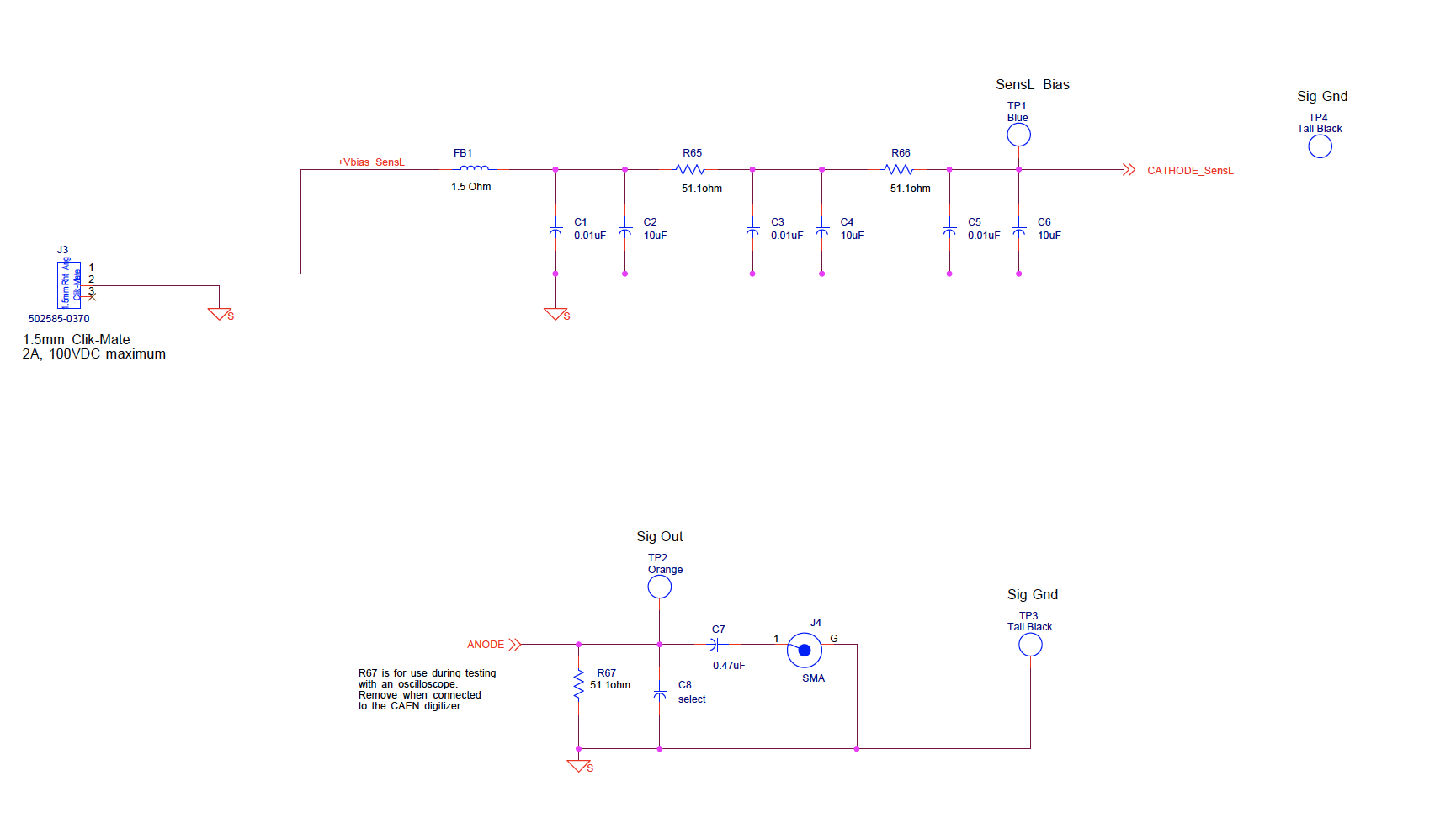}
			\caption{}
		\end{subfigure}	
		\begin{subfigure}[h]{\columnwidth} 
			\includegraphics[width=\columnwidth]{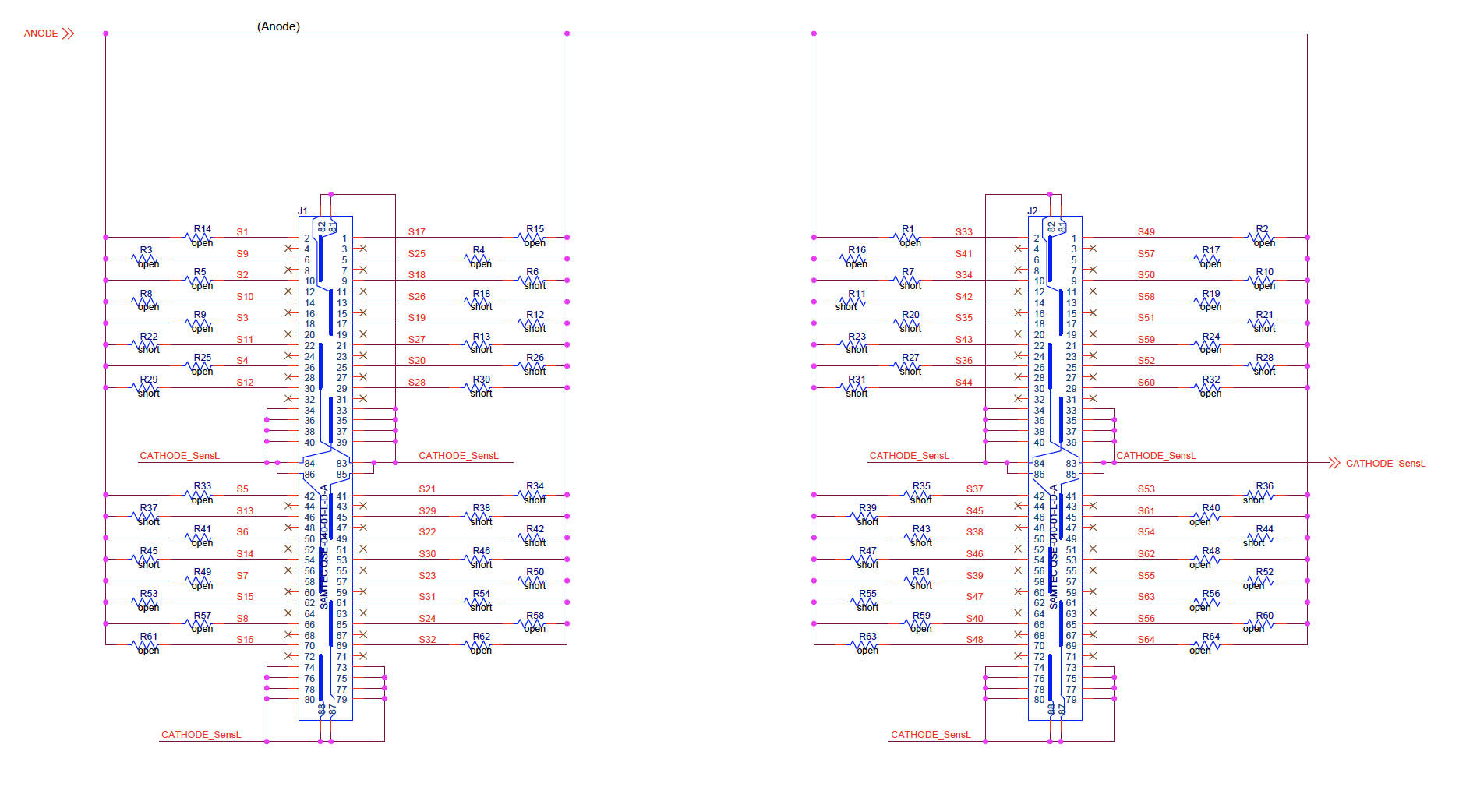}
			\caption{}
		\end{subfigure}	
	\caption{The top (a) and bottom (b) side components of the passive summing board.}
	\label{fig:fig1} 
\end{figure}

\begin{figure}[!htbp]
		\centering
			\includegraphics[angle=90,width=0.49\columnwidth]{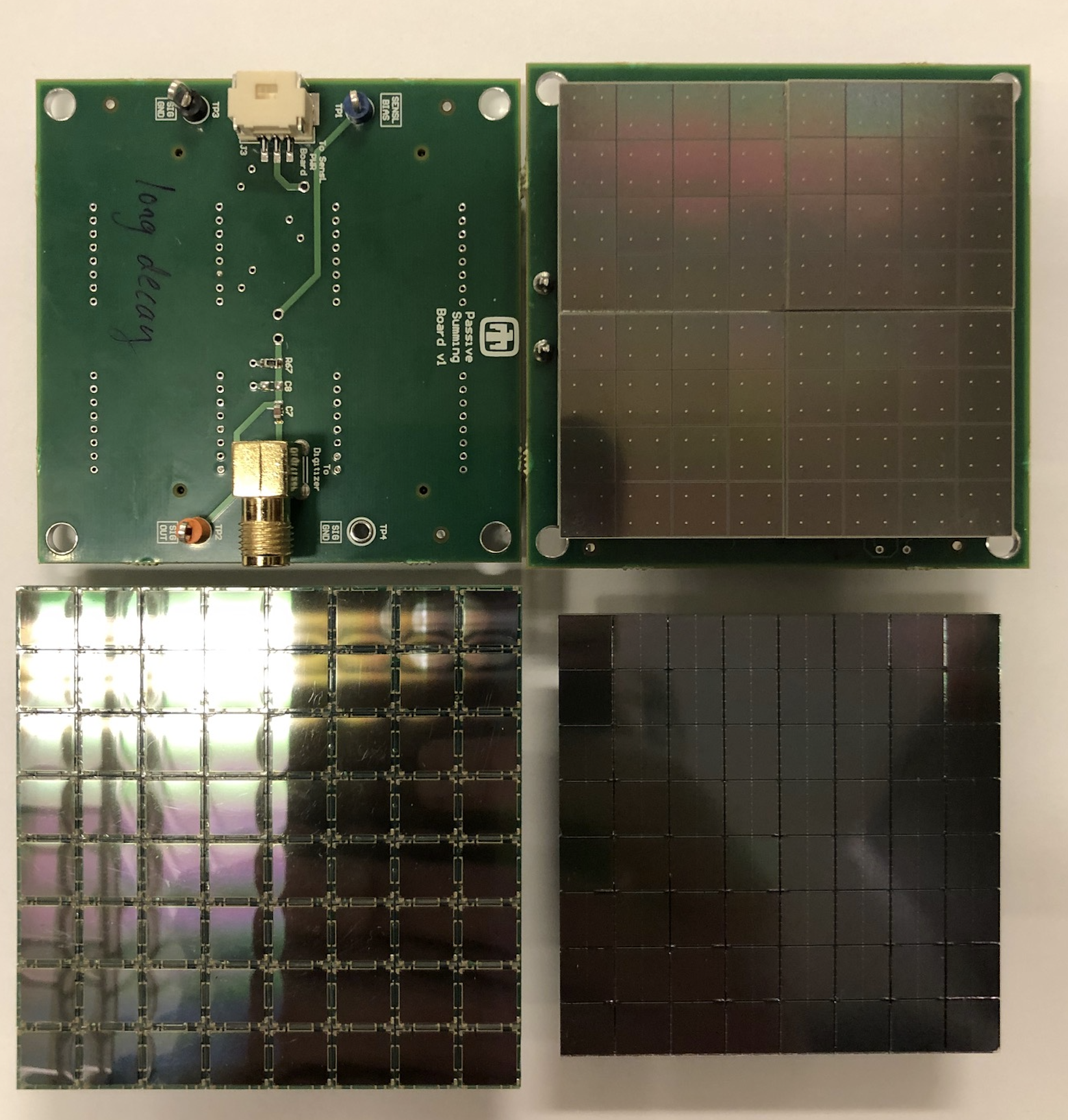}
			\caption{Clockwise from the top left: the array of Hamamatsu S13361-6050 MPPCs assembled in an 8x8 array, SensL's ArrayJ-60035-64P-PCB, SensL's ArrayC-60035-64P-PCB, and the passive summing board developed for this work.}
	\label{fig:fig2} 
\end{figure}

In order to realize the complete replacement of PMT assemblies with SiPMs, we developed a compact printed circuit board (PCB) that mates to the commercial arrays and sums the individual responses of each SiPM in the array. While SensL provides a PCB (ArrayX-BOB6\_64S) that mates to their ArrayC-60035-64P-PCB and ArrayJ-60035-64P-PCB, it is nearly twice the cross section of the SiPM arrays. In addition, the board requires the user to assemble the readout circuit with lead-wires provided by the manufacturer. While this allows the board to be adaptable to different user needs, such as changing the capacitor in the readout circuit for different decay times or reading out individual pixels in addition to the sum, the lead-wires cause the board to be susceptible to noise pickup. Early testing with the ArrayX-BOB6\_64S board indicated that, although it does not come in an ideal form factor and required external shielding, the pulse-shape and timing performance was comparable to photo-multiplier readout: Appendix A summaries these results. Based on these tests, we designed a summing board much like SensL's ArrayX-BOB6\_64S in which all pixels are passively summed into one readout circuit, but with a smaller physical cross section. The circuit diagram for our board is shown in Figure \ref{fig:fig1}. The boards are the same size as the C-series array (5.7x5.7 mm$^2$), and mate directly to the connectors on the back of the arrays, with the exception of the Hamamatsu array. All the arrays along with the summing board are shown in Figure \ref{fig:fig2}.

The SiPM array is configured for a positive bias voltage which is applied to the SiPM cathodes via the signal labeled CATHODE\_SensL.  The cathodes of all the elements of the array are connected together.  The output signal is extracted from the SensL Standard Output (SOUT) terminal.  All the standard outputs are summed into the signal labeled ANODE through a resistor network whose purpose is explained below. The technique for readout was derived from the circuits shown in SensL application note AND9782/D. 

The top side of the board provides filtering of the bias voltage, impedance matching and filtering of the output pulses. The connection to the bias voltage for the array is J3. A Ferrite Bead (FB1) provides high frequency noise rejection from the bench power supplies used to generate the bias voltage. C1, C2, R65, C3, C4, R66, C5 and C6 form a two-stage R-C low pass filter.  The filter is a modification of the SensL's, outlined in their application note AND9782/D.  Additional filtering was implemented to provide better noise reduction. R67, C8, C7 and J4 form the output stage.  R67 provides impedance matching when interfacing to high impedance readout devices such as an oscilloscope.  It is removed when interfacing to 50 ohm input impedance devices.  C8 is a select value capacitor that is used to suppress high frequency noise.  Its value will typically be open, or at most, a few picofarads.  C7 is the DC blocking capacitor that removes any DC offset voltage present.  It also has an effect on the shape the ``tail" of the output pulse.  Smaller values will produce faster decay times.

On the bottom side of the board, J1 and J2 are SAMTEC part number QSE-040-01-F-D-A. These connectors mate to the connectors on the SensL SiPM 64 element array board. For these measurements, only the SiPM SOUT signals were summed.  The Fast Outputs (FOUT) are left unconnected per SensL's recommendation. R1 through R64 form the passive summing network.  Their values are limited to ``open" (no resistor installed), or ``short" (a zero ohm resistor installed).  Their purpose is to allow the user to select which SiPM pixels are summed into the final output signal.  As depicted, the network is configured to enable only those pixels that are illuminated by a 5 cm diameter scintillator. Equal length traces are used for all these connections to eliminate differences in the time delay of the rise and decay of the summed output signals.
 
\section{Experimental Setup}
Each array was biased with a BK Precision 1761 DC power supply. For the Hamamatsu array, we used a bias of 56 V as recommended by the manufacturer by connecting two channels of the power supply in series. For the SensL SiPMs, we operate at the highest over-voltage recommended by the manufacturer, or 30 V for the J-series and 29.5 V for the C-series. This value is motivated by a desire for the best timing resolution \cite{catteneo}. The manufacturer also reports increasing photo-detection efficiency as a function of over-voltage. 

For the 5 cm diameter crystal measurements, the crystal is wrapped with Teflon on all surfaces that were not coupled to the SiPM array. Optical grease is used to couple the readout surface of the crystal to the SiPM array. The crystal, SiPM array, and summing board circuit were placed inside a light-tight enclosure. The summed output from each array was digitized with CAEN's DT5730 500 MHz, 14-bit digitizer, with the dynamic range set to 500 mV. For measurements with multiple digitizer inputs, the digitizer was asynchronously triggering on all channels at a threshold well below the 341 keVee Compton edge of 511 keV gammas. We acquired 2400 ns for each trace in order to capture the entire SiPM waveform, with 200 ns in the pre-trigger window.

\section{Readout Characterizations}
\begin{figure}[!htbp]
	\begin{subfigure}[h]{0.49\columnwidth} 
		\centering
		\includegraphics[width=\columnwidth]{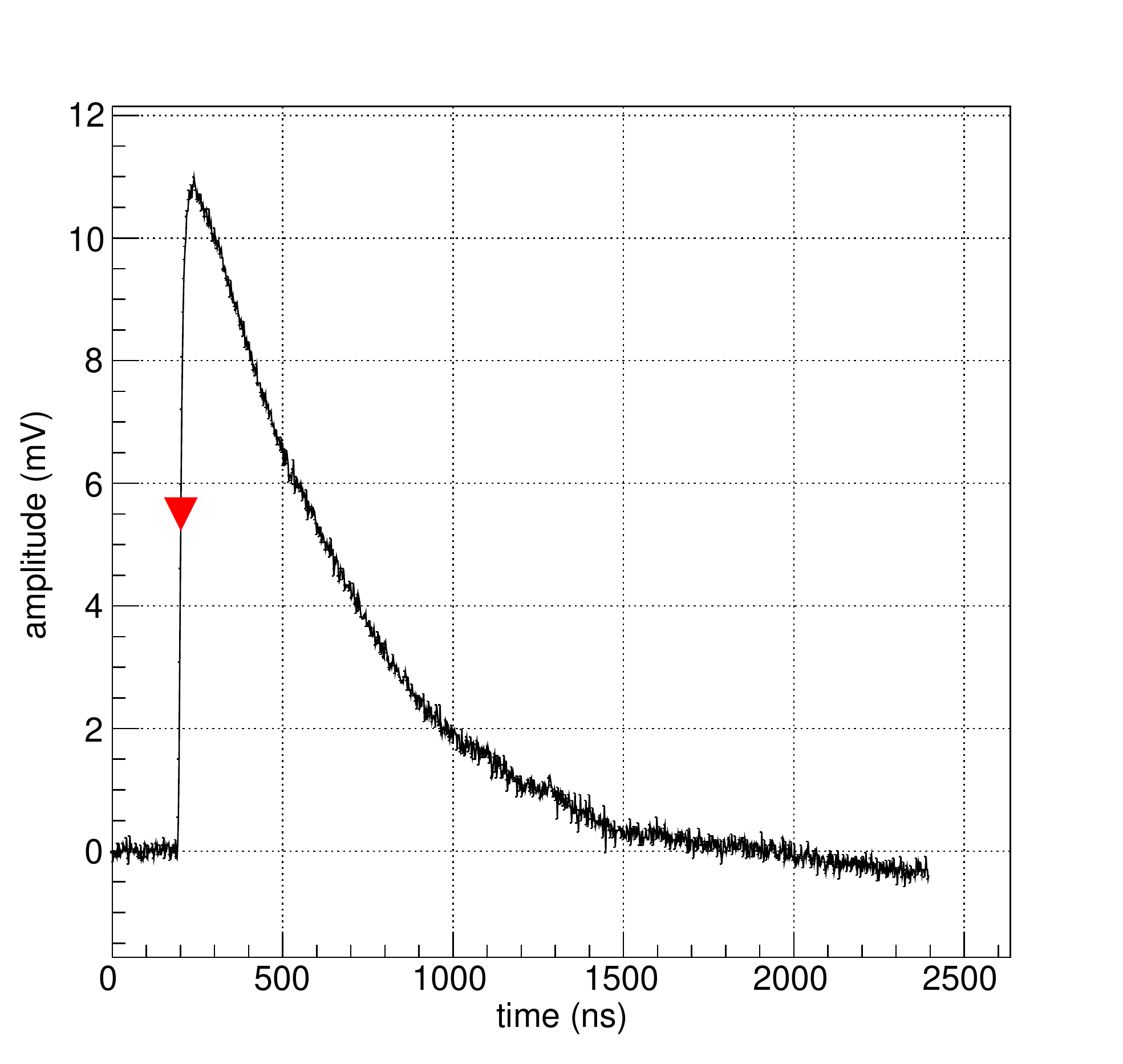}
		\caption{}
	\end{subfigure}	
	\begin{subfigure}[h]{0.49\columnwidth} 
		\centering
		\includegraphics[width=\columnwidth]{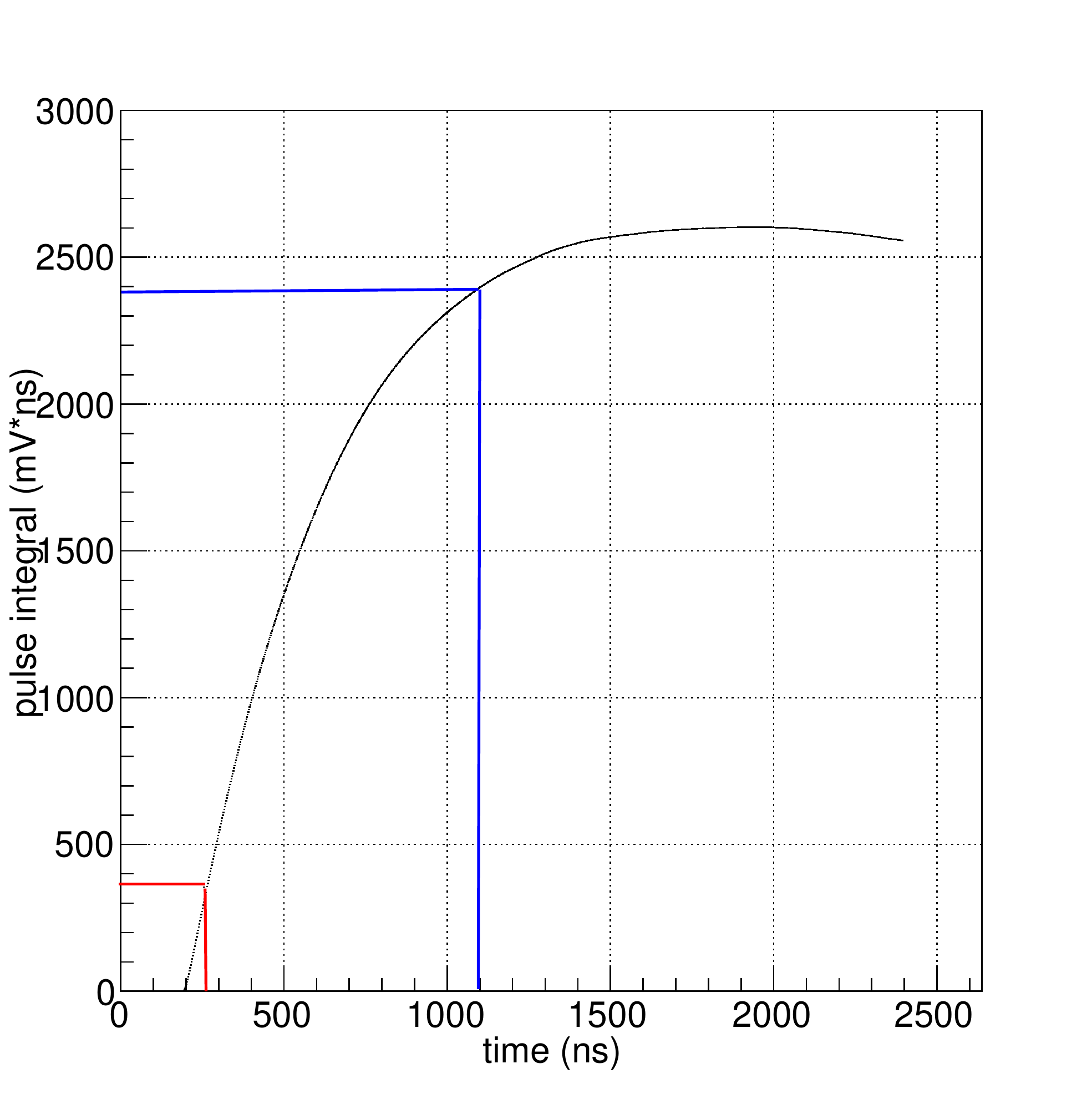}
		\caption{}
	\end{subfigure}
\begin{subfigure}[h]{0.49\columnwidth} 
		\centering
		\includegraphics[width=\columnwidth]{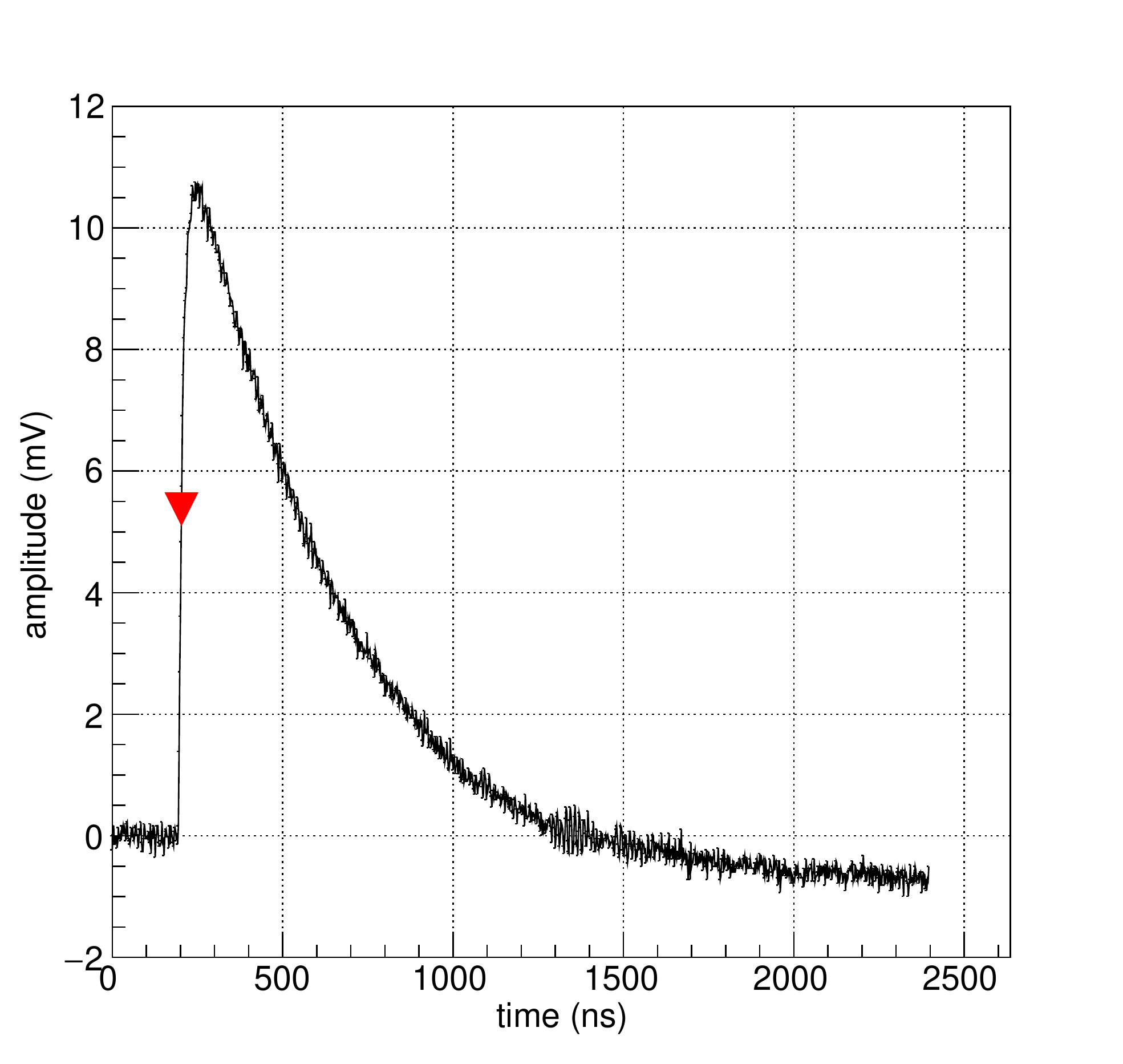}
		\caption{}
	\end{subfigure}	
	\begin{subfigure}[h]{0.49\columnwidth} 
		\centering
		\includegraphics[width=\columnwidth]{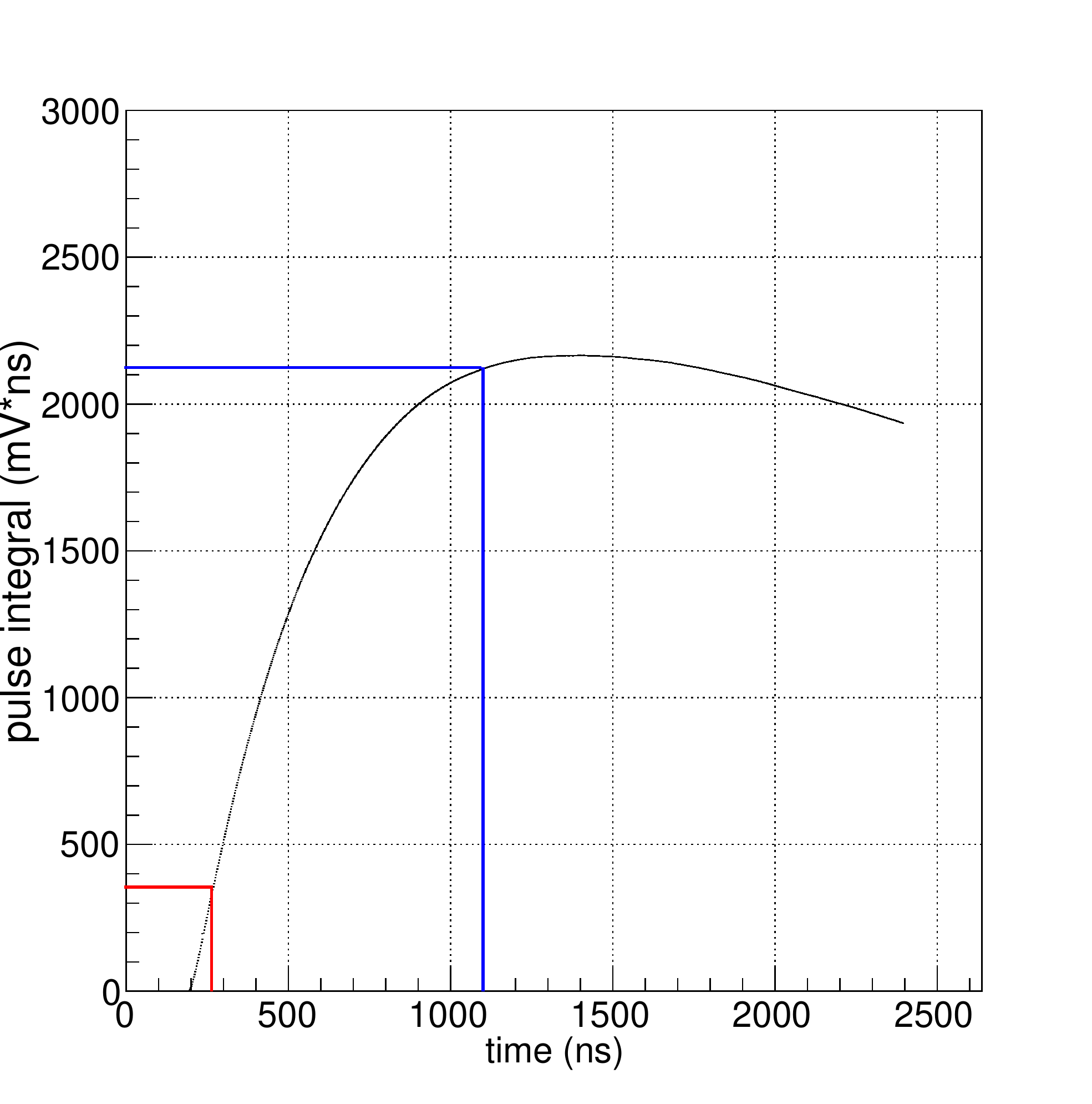}
		\caption{}
	\end{subfigure}
\caption{Example summed traces from SensL's ArrayJ-60035-64P-PCB (a) and the Hamamatsu S13361-6050 array (c), with the pulse time indicated by the red arrow. Example cumulative distributions used to calculate energy windows for the corresponding traces are shown for SensL's ArrayJ-60035-64P-PCB (b) and the Hamamatsu S13361-6050 array (d). The prompt window and corresponding value is shown in red, and the total window and corresponding value is shown in blue. }
\label{fig:fig3} 
\end{figure}

All analysis operations are written in C++ utilizing algorithms from the ROOT analysis toolkit \cite{root}. Before energy, timing, and PSD characterizations, the waveforms are baseline subtracted using 
the first 120 ns in the trace.   The maximum sample value is used to measure the total energy deposition. The pulse time is the linearly interpolated value between samples corresponding to 50\% 
of the maximum value on the rising edge of the pulse. Figure \ref{fig:fig3} shows example waveforms from the J-series (a) and Hamamatsu (c) arrays, with the pulse time indicated by the red arrow. 
Finally, for pulse shape measurements, we construct the cumulative distribution of the pulse and linearly-interpolate between samples for the value corresponding to the desired integration window. 
Figure \ref{fig:fig3} shows examples of the cumulative distribution for the J-series (b) and Hamamatsu (d) arrays. The corresponding values for the prompt and total energy window are indicated in red 
and blue: the total energy window was chosen to be well outside the point where the pulse undershoots below the baseline.

\subsection{Energy Resolution}
\begin{figure}[!htbp]
	\begin{subfigure}[h]{0.32\columnwidth} 
		\centering
		\includegraphics[width=\columnwidth]{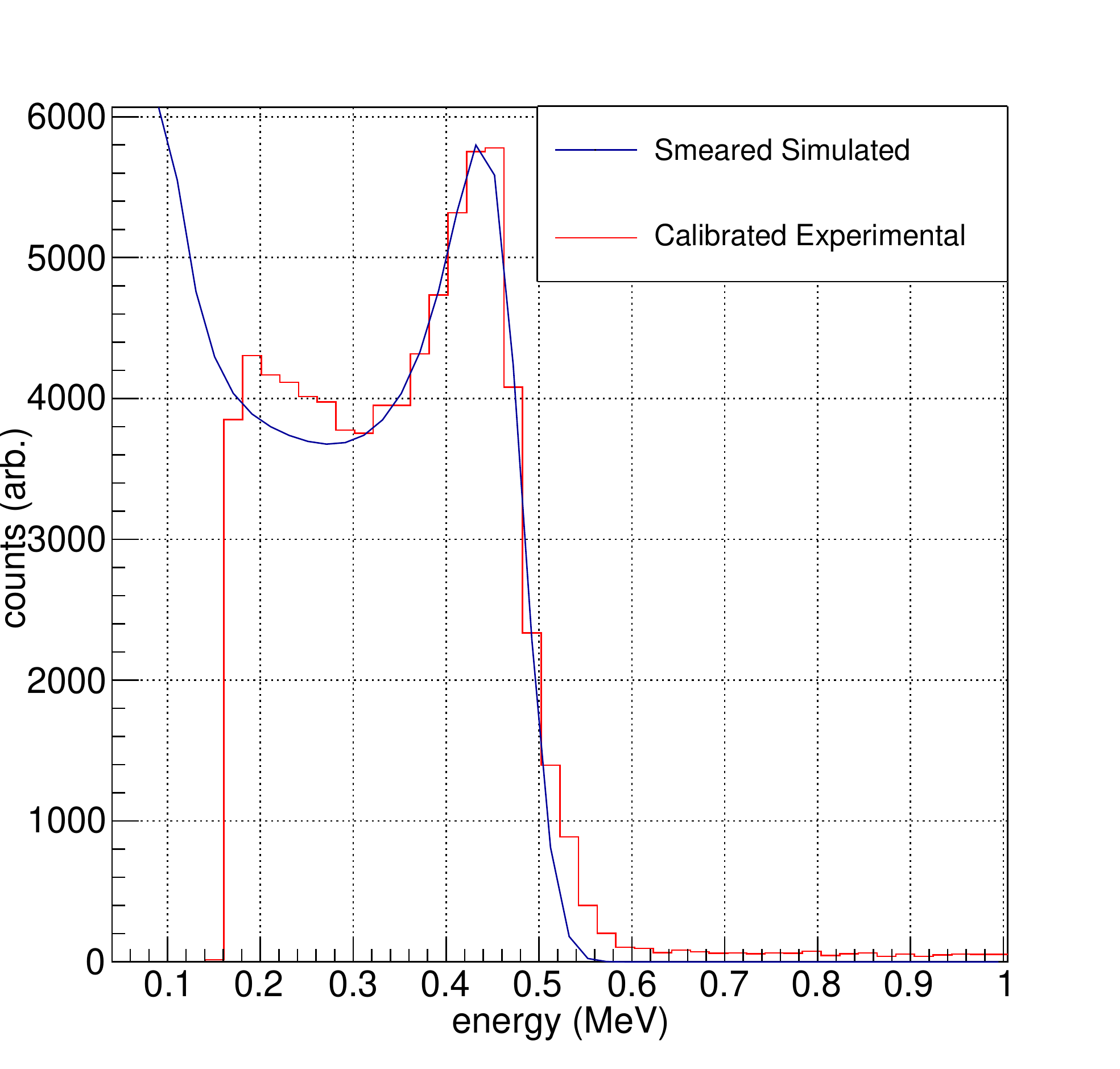}
		\caption{}
	\end{subfigure}	
	\begin{subfigure}[h]{0.32\columnwidth} 
		\centering
		\includegraphics[width=\columnwidth]{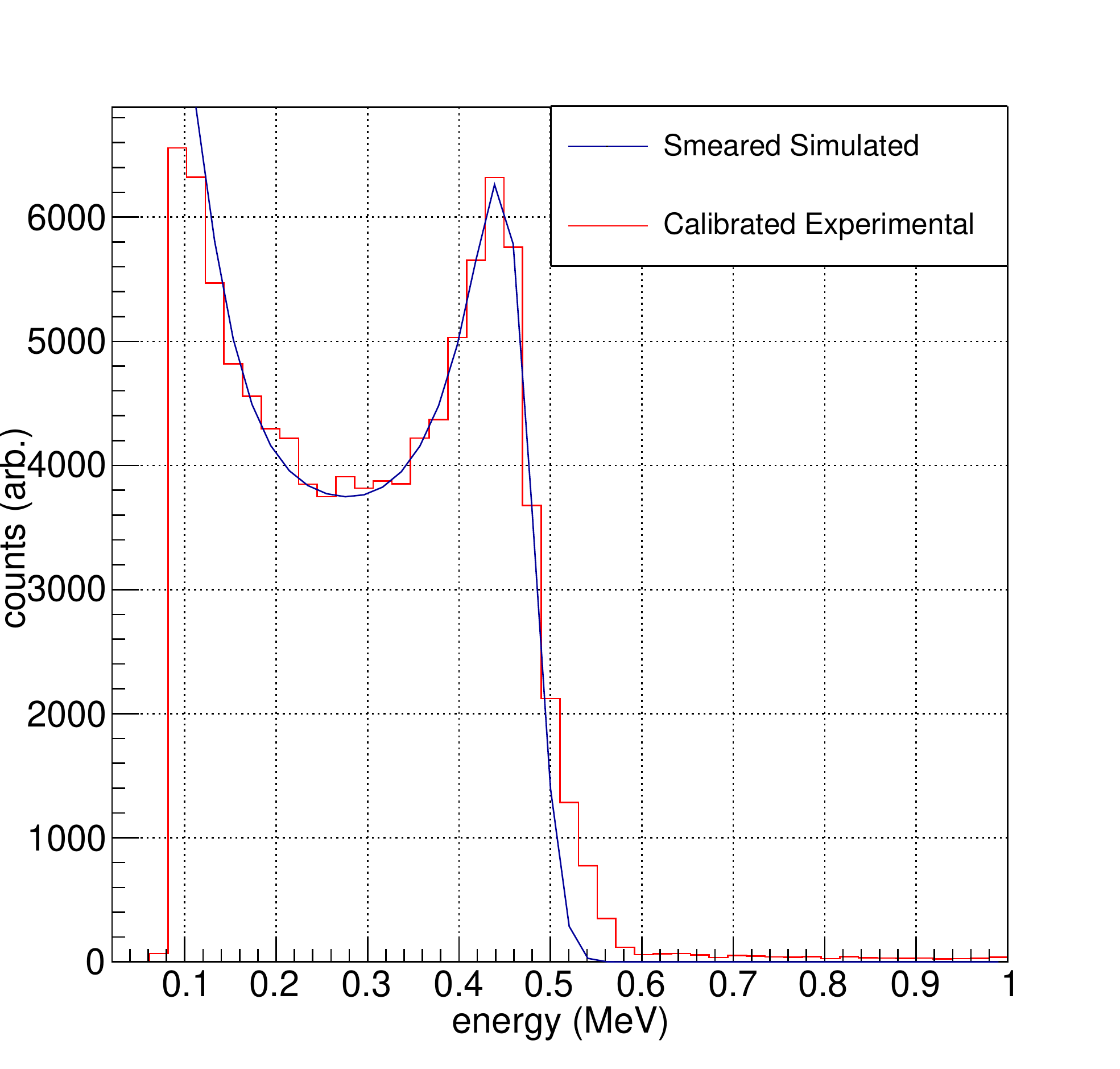}
		\caption{}
	\end{subfigure}
	\begin{subfigure}[h]{0.32\columnwidth} 
		\centering
		\includegraphics[width=\columnwidth]{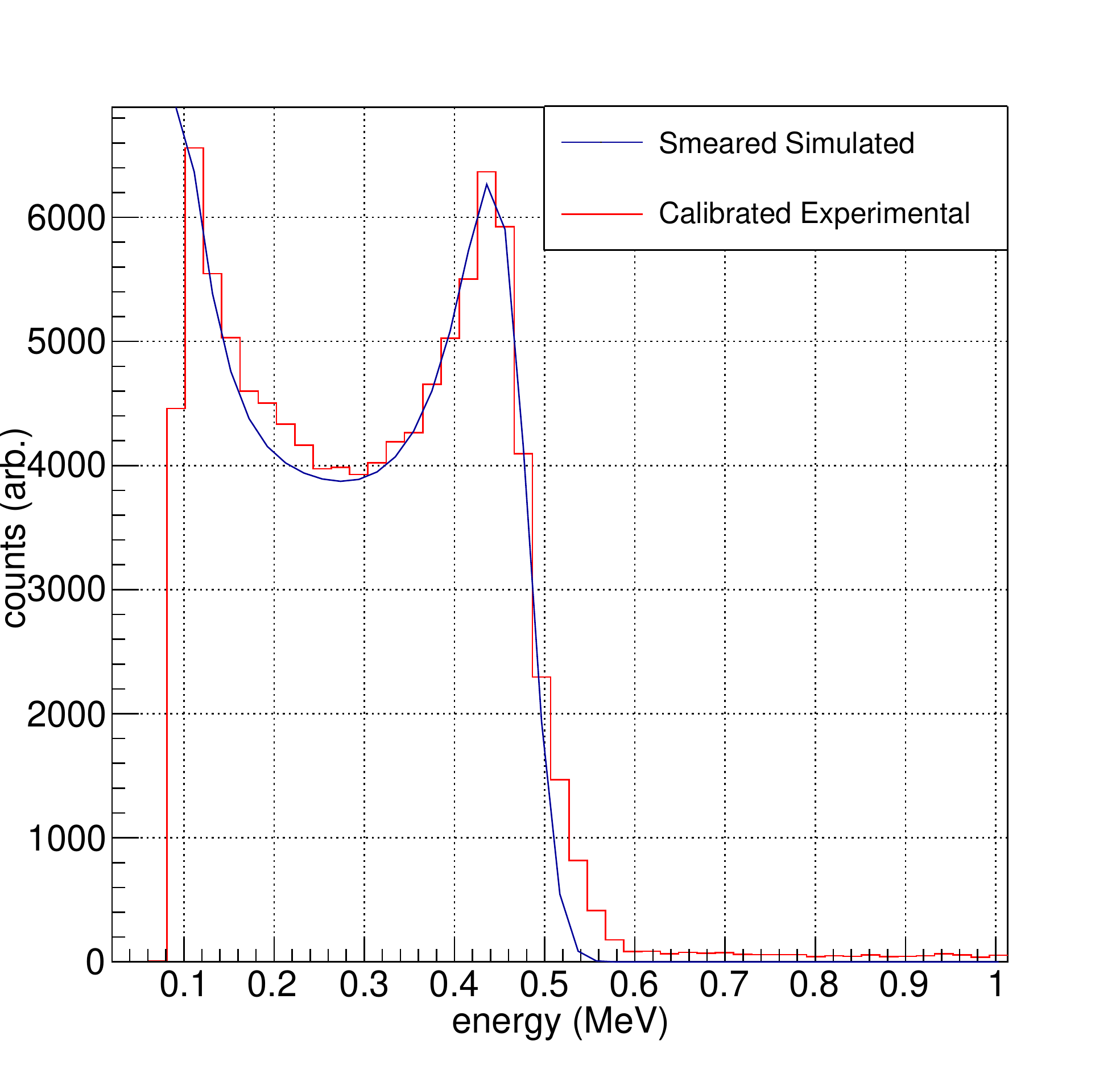}
		\caption{}
	\end{subfigure}

\caption{The calibrated energy spectrum measured from a $^{137}$Cs source (red) and the smeared Klein--Nishina prediction (blue) for SensL's ArrayC-60035-64P-PCB (a), SensL's ArrayJ-60035-64P-PCB (b), and the Hamamatsu S13361-6050 array (c).}
\label{fig:fig4} 
\end{figure}

\begin{figure}[!htbp]
	\begin{subfigure}[h]{0.49\columnwidth} 
		\centering
		\includegraphics[width=\columnwidth]{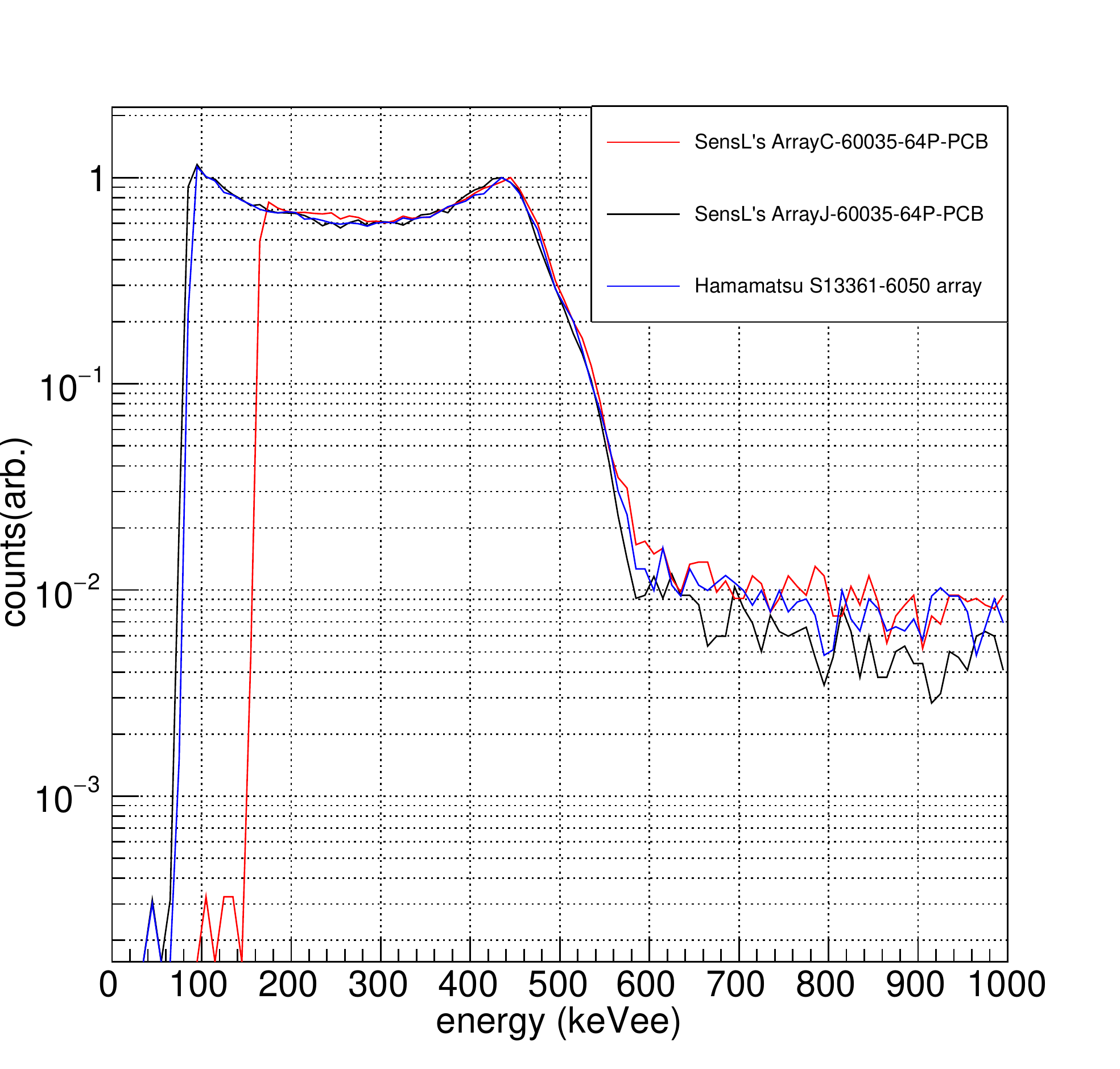}
		\caption{}
	\end{subfigure}	
	\begin{subfigure}[h]{0.49\columnwidth} 
		\centering
		\includegraphics[width=\columnwidth]{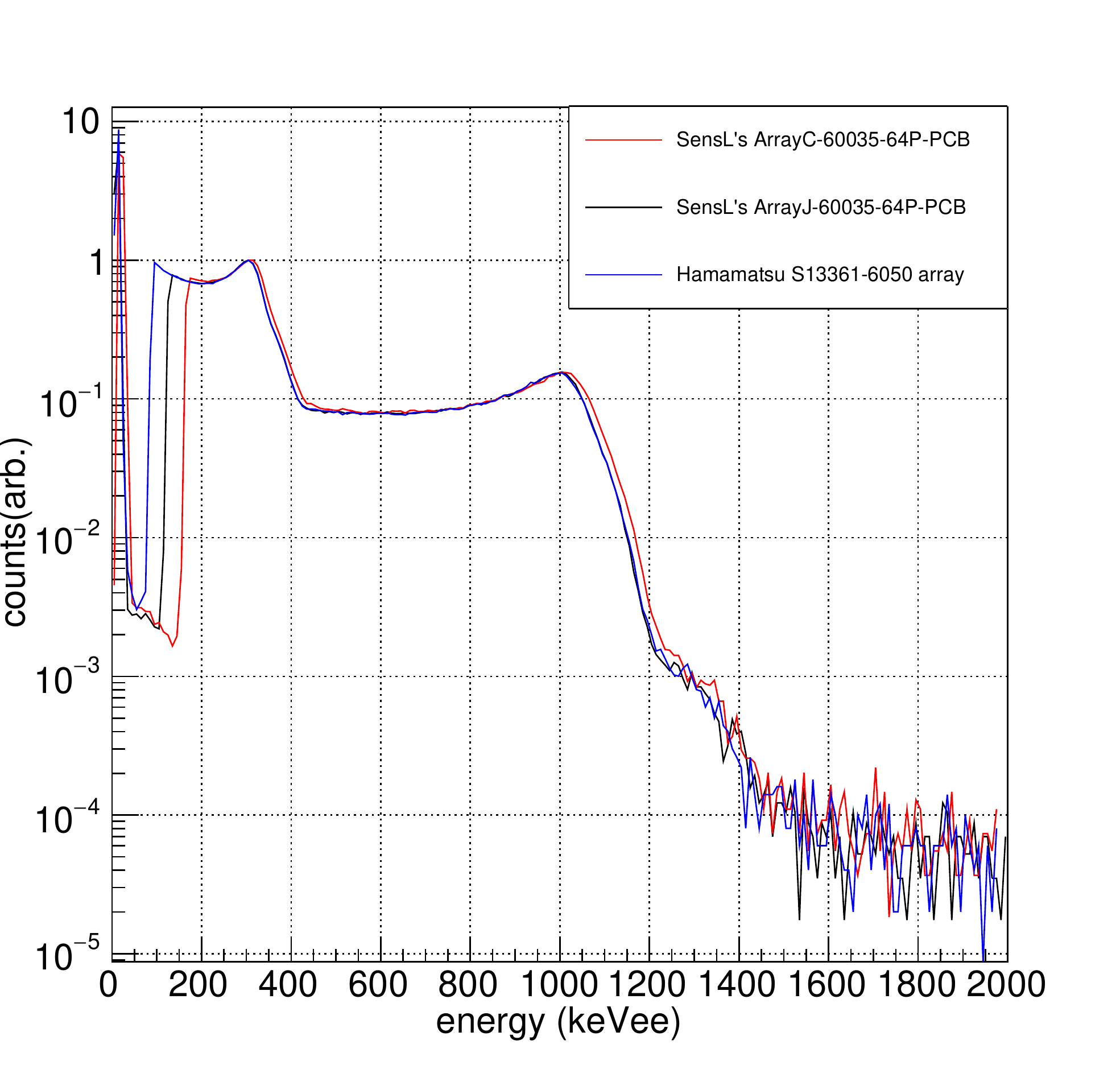}
		\caption{}
	\end{subfigure}
\caption{The calibrated energy spectrum measured from a $^{137}$Cs source (a) and a $^{22}$Na source (b).}
\label{fig:fig5} 
\end{figure}

The energy resolution is measured by fitting the measured energy spectrum to the Klein--Nishina spectrum.  The energy spectrum from $^{137}$Cs and $^{22}$Na are used 
to compare to the expected spectrum calculated from the Klein--Nishina prediction. First, the experimental result is converted to keVee units with a linear 
transformation:
\begin{linenomath*}
\begin{equation}
E_{keVee} = q_0E_{mV}+q_1.
\end{equation}
\end{linenomath*}
Next, an arbitrary y-scale, $q_2$, is applied to the Klein-Nishina prediction for the gamma emissions, and a gaussian 
convolution is performed with a standard deviation of:
\begin{linenomath*}
\begin{equation}
\frac{\sigma}{E_{keVee}}= \sqrt{q_3^2 + \frac{q_4^2}{E_{keVee}} + \frac{q_5^2}{E_{keVee}^2} }.
\label{eq:res}
\end{equation}
\end{linenomath*}
The $^{137}$Cs energy spectrum fit results for each array are shown in Figure \ref{fig:fig4}, with the measured spectrum in red and the smeared Klein--Nishina prediction in blue. Figure \ref{fig:fig5} show the measured spectra for all three arrays from $^{137}$Cs (a) and $^{22}$Na (b). The values and errors in Table \ref{tab:tab1} are the mean and standard deviation of the results of fits to 10 bootstrapped energy spectra obtained by sampling the energy spectra with replacement. We do not measure a significant difference in energy resolution between the different arrays, possibly indicating that the intrinsic resolution of the scintillator along with the random processes involved in light transport are the dominant factors. 

\begin{table}
\centering
\begin{tabular}{|c||c|c|c|}
\hline
Array					&341 keVee (\%)	& 478 keVee (\%)		& 1057 keVee	(\%)		\\
\hline
ArrayC-60035-64P-PCB		&13.5$\pm$1.0		&14.6$\pm$0.7		&8.2$\pm$0.6			\\
ArrayJ-60035-64P-PCB		&13.6$\pm$1.8		&13.8$\pm$0.3		&7.8$\pm$0.9			\\
Hamamatsu S13361-6050		&13.7$\pm$0.9 	&14.7$\pm$0.4		&8.2$\pm$0.7			\\
\hline
\end{tabular}
\caption{The energy resolution ($\sigma_{E}$) for each array for three different energies. The values in the first row are the Compton edges of the 511 and 1200 keV lines from $^{22}$Na and the 662 keV line from $^{137}$Cs. The errors are the standard deviation of the results of fits to 10 bootstrapped energy spectra obtained by sampling the energy spectra with replacement.}
\label{tab:tab1}
\end{table}

\subsection{Timing Resolution}
\begin{figure}[!htbp]
		\centering
		\includegraphics[width=\columnwidth]{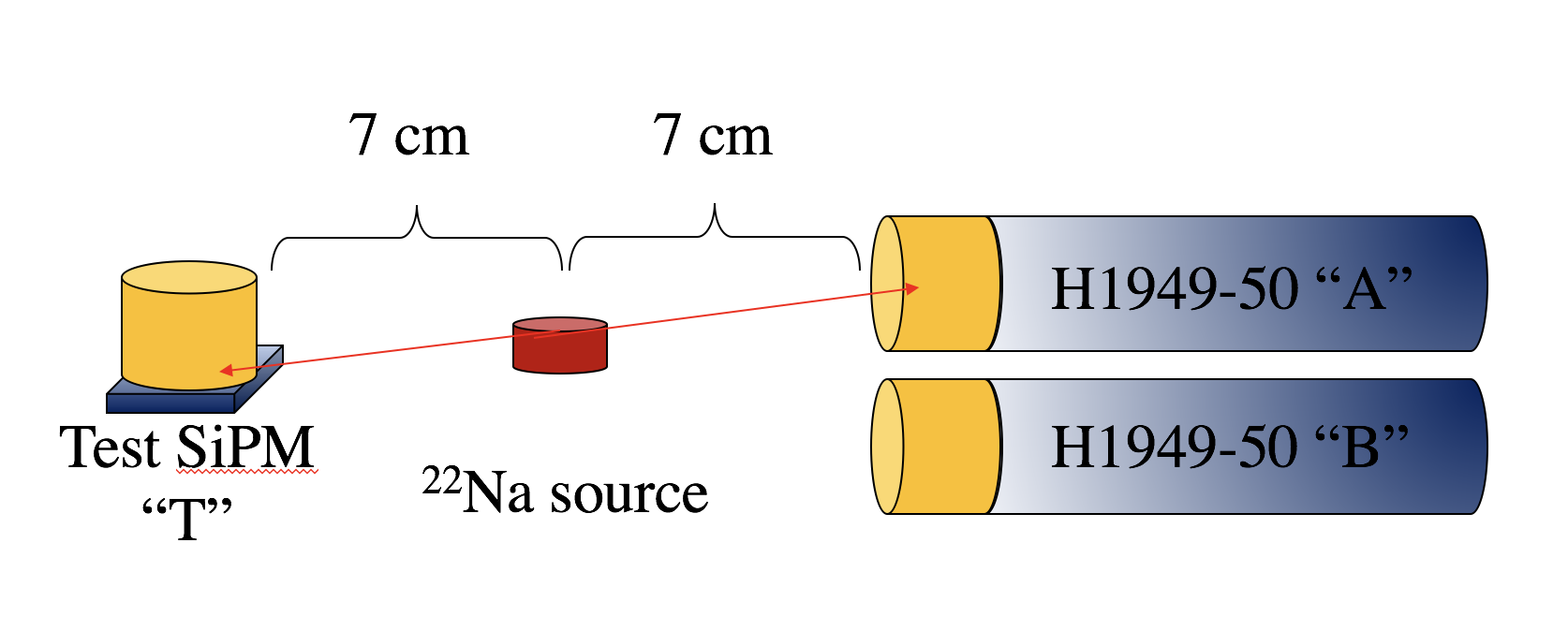}
\caption{The timing response measurement setup. A $^{22}$Na source is placed between the crystal readout with the test SiPM array, denoted ``T", and two additional stilbene crystals readout with H1940-50 PMTs, denoted ``A" and ``B". A 511 keV gamma measured in ``T" can result in a coincident 511 keV interaction in ``A" or ``B": an example is shown in red.}
\label{fig:fig6} 
\end{figure}

\begin{figure}[!htbp]
	\begin{subfigure}[h]{0.32\columnwidth} 
		\centering
		\includegraphics[width=\columnwidth]{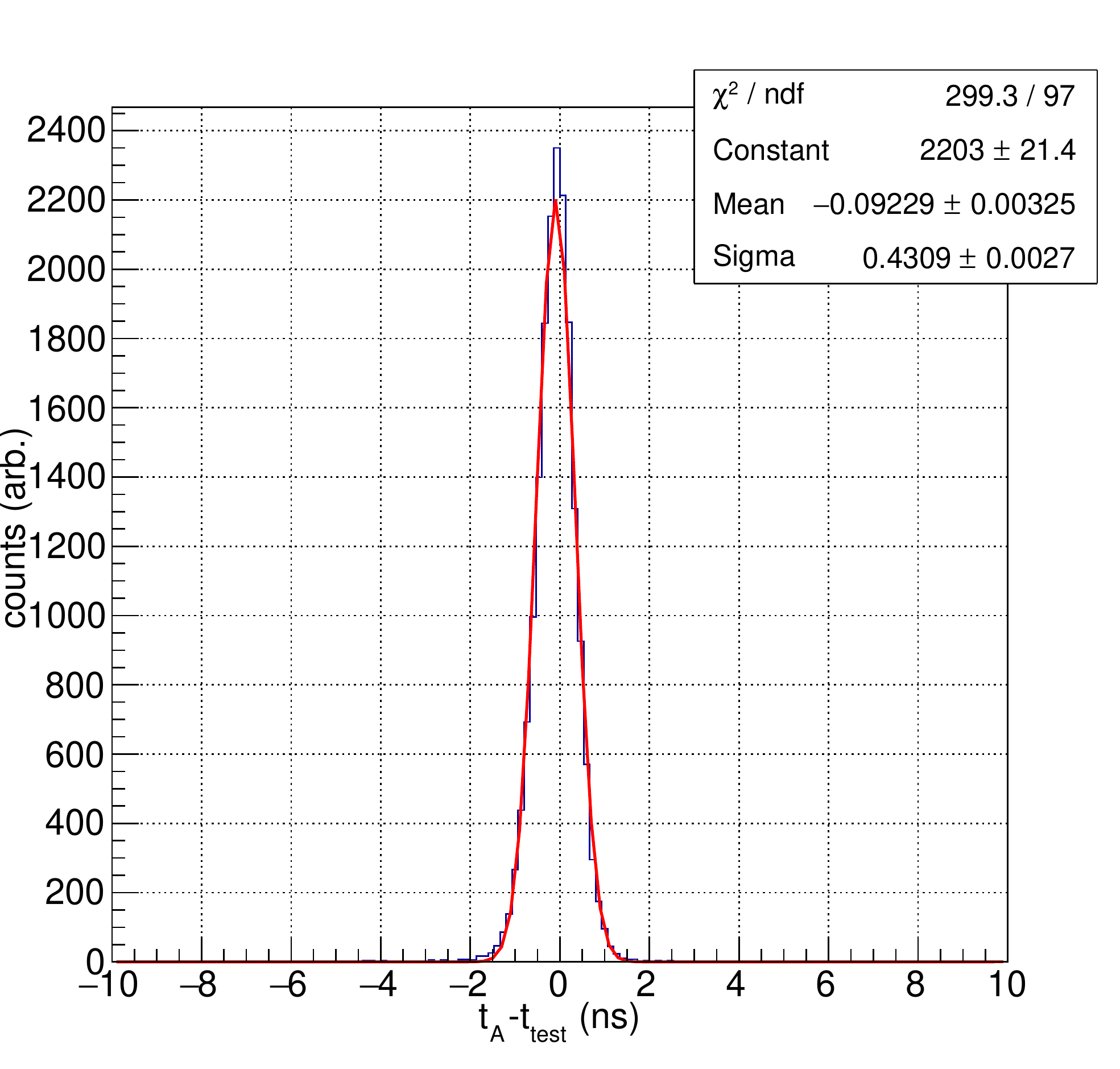}
		\caption{}
	\end{subfigure}	
	\begin{subfigure}[h]{0.32\columnwidth} 
		\centering
		\includegraphics[width=\columnwidth]{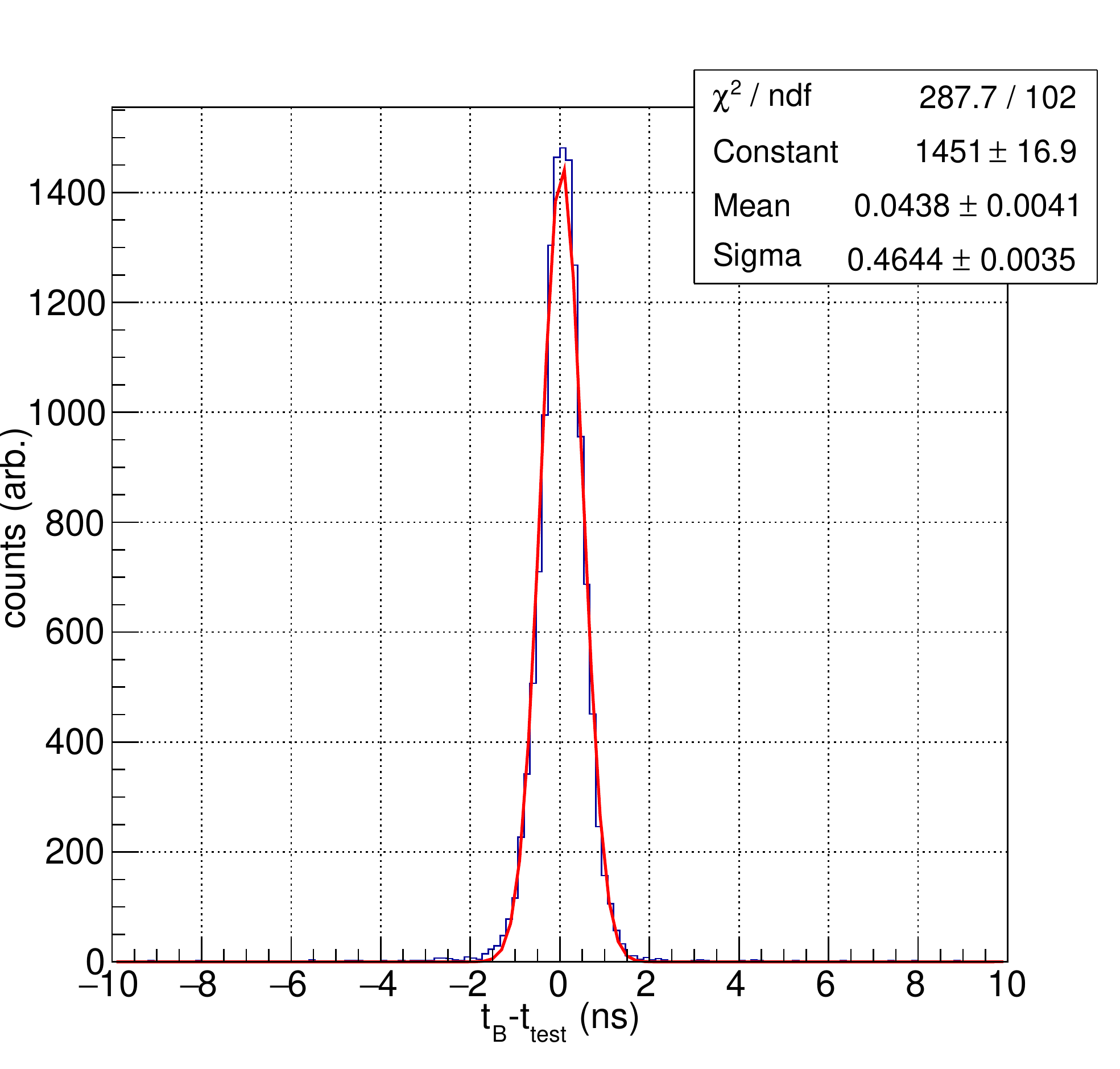}
		\caption{}
	\end{subfigure}
	\begin{subfigure}[h]{0.32\columnwidth} 
		\centering
		\includegraphics[width=\columnwidth]{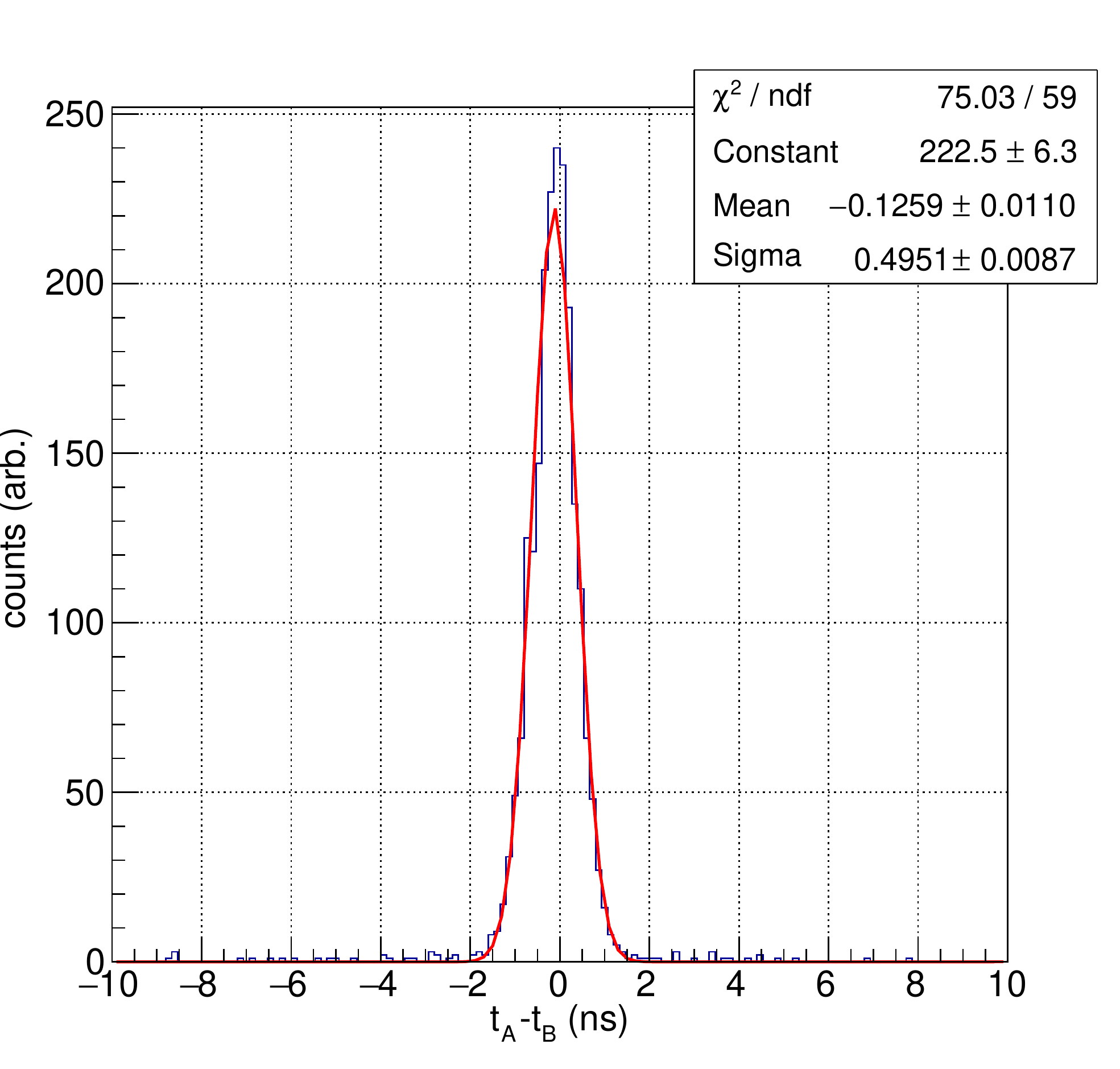}
		\caption{}
	\end{subfigure}

\caption{The timing distributions and fits for the ArrayJ-60035-64P-PCB: (a) $t_A-t_{T}$, (b) $t_B-t_{T}$, and (c) $t_A-t_{B}$. The errors in the table are statistical on the parameters.}
\label{fig:fig7} 
\end{figure}

The timing response of the SiPM array readout is measured by the coincident interactions of the two back-to-back simultaneously emitted 511 keV gammas resulting from positron annihilation in a $^{22}$Na source. Figure 
\ref{fig:fig6} is a pictorial representation of the measurement. Two 5 cm diameter stilbene crystals read out with H1949-50 Hamamatsu PMTs, denoted  ``A" and ``B", are placed opposite and equidistant from the 5 cm diameter
stilbene readout with the SiPM array, denoted ``T".  The $^{22}$Na is placed in the center. In this configuration, one gamma interacting in ``T" can result in the second gamma interacting at the same time in either 
``A" or ``B". The timing resolution of the SiPM array readout can then be determined by a combination of timing differences between the three: 
\begin{linenomath*}
\begin{eqnarray}
\sigma_{A-T}^2 &=& \sigma_{T}^2 + \sigma_{A}^2 \\
\sigma_{B-T}^2 &=& \sigma_{T}^2 + \sigma_{B}^2 \\
\sigma_{A-B}^2 &=& \sigma_{A}^2 + \sigma_{B}^2.
\end{eqnarray}
\end{linenomath*}
Solving for $\sigma_T$ gives:
\begin{linenomath*}
\begin{equation}
\sigma_T = \sqrt{\frac{\sigma_{A-T}^2 + \sigma_{B-T}^2 - \sigma_{A-B}^2}{2}}.
\label{eq:1}
\end{equation}
\end{linenomath*}
The distributions of timing differences (e.g. $t_A-t_T$) are fit to a Gaussian function to determine the standard deviations (e.g. $\sigma_{A-T}$). An example set of distributions is shown in 
Figure \ref{fig:fig7} for the J-series array. Events in these distributions are limited to a 100-400 keVee energy window to assure only contributions from the 511 keV gammas. The standard deviations 
for each configuration are tabulated in Table \ref{tab:tab2}. Due to the size of the crystal, there is potential for a significant contribution of the timing resolution to be from variations in interaction locations in either crystal. A Geant4 \cite{geant4} simulation of the experimental setup indicates that the majority of interactions occur near the entry surface for a particular trajectory, leading to variations on the order of tens of ps. Light propagation will contribute an additional factor. What our measurements present here is not the coincident timing attributed only to the readout, but the coincident timing resolution of this particular detector geometry and readout.
To estimate the error on the timing distributions, we assume that the interaction location distributions are similar for each dataset so that 
$\sigma_{A-B}$ for each configuration can be treated as the same quantity: the standard deviation of all three $\sigma_{A-B}$ measurements is 23 ps. The standard deviation of $\sigma_{A-T}$ ($\sigma_{B-T}$) is 
dependent on both $\sigma_A$ ($\sigma_B$) and $\sigma_T$, and therefore cannot be calculated with this data. We therefore combine the measurements of $\sigma_{A-T}$ and $\sigma_{B-T}$ to estimate their errors: this amounts to an incorrect assumption of $\sigma_A=\sigma_B$, however the degree to which this assumption is true is contained within the standard deviation between the measurements. Our data indicate that the J-series outperforms the Hamamatsu in timing resolution by a significant factor. It may be that the adaptor board for the Hamamatsu array is a contributing factor.

\begin{table}
\centering
\begin{tabular}{|c||c|c|c||c|}
\hline
Array					&$\sigma_{A-T}$ (ps)&$\sigma_{B-T}$ (ps)	&$\sigma_{A-B}$ (ps) 	&$\sigma_{T}$ (ps)\\
\hline
ArrayC-60035-64P-PCB		&551$\pm$26		&588$\pm$26			&454$\pm$23			&471$\pm$25	\\
ArrayJ-60035-64P-PCB		&429$\pm$24		&463$\pm$24			&495$\pm$23			&277$\pm$34	\\
Hamamatsu S13361-6050		&523$\pm$32		&568$\pm$32			&493$\pm$23			&420$\pm$32	\\
\hline
\end{tabular}
\caption{The standard deviations of each timing difference distribution, and the resulting timing resolution for each array. The energy window for events is 100-400 keVee. The errors are calculated assuming that $\sigma_A=\sigma_B$ for each configuration, and that for the three different configurations $\sigma_{A-B}$ is the same.}
\label{tab:tab2}
\end{table}

\subsection{Particle Identification}
\begin{figure}[!htbp]
	\begin{subfigure}[h]{0.32\columnwidth} 
		\centering
		\includegraphics[width=\columnwidth]{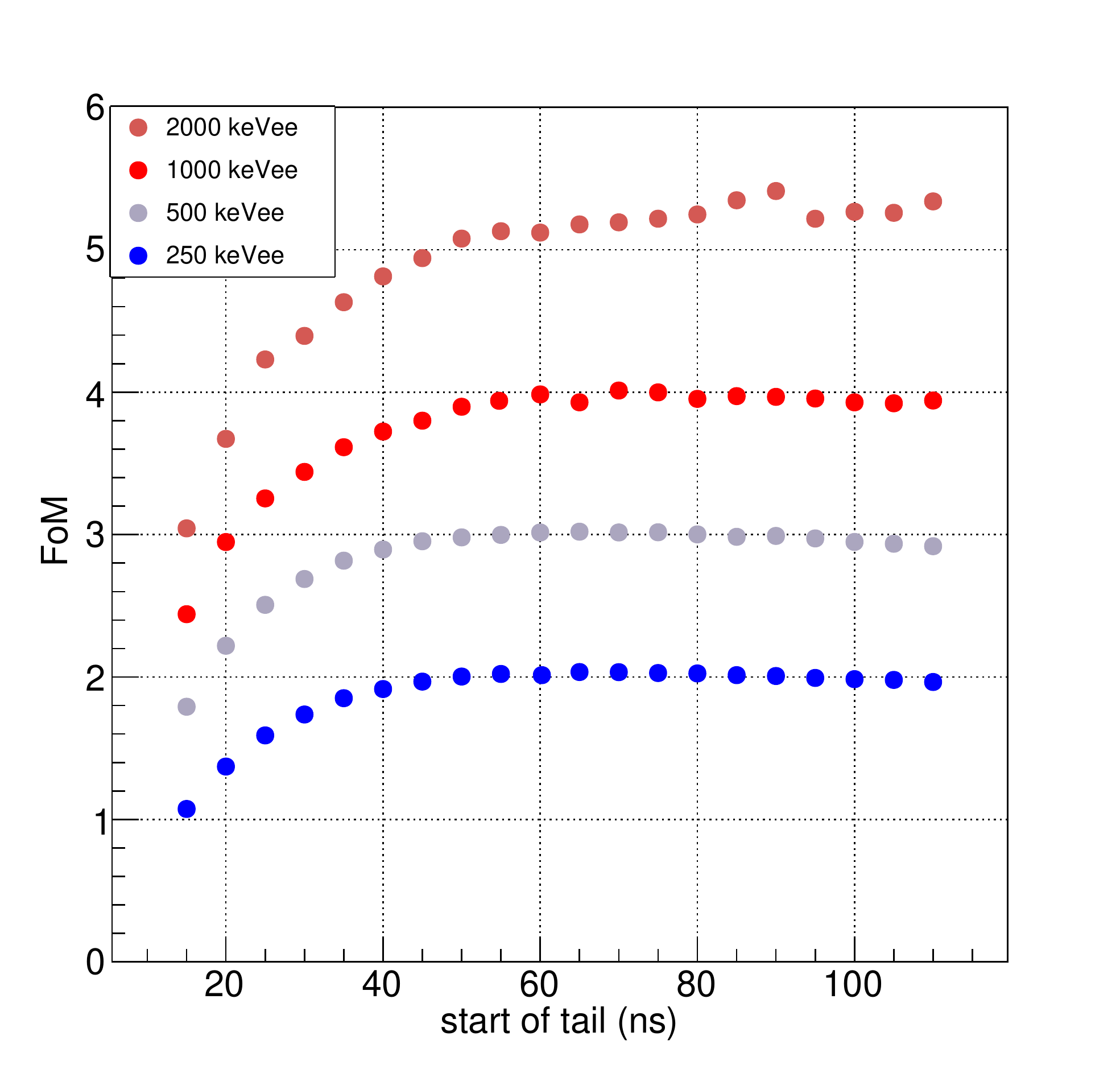}
		\caption{}
	\end{subfigure}	
	\begin{subfigure}[h]{0.32\columnwidth} 
		\centering
		\includegraphics[width=\columnwidth]{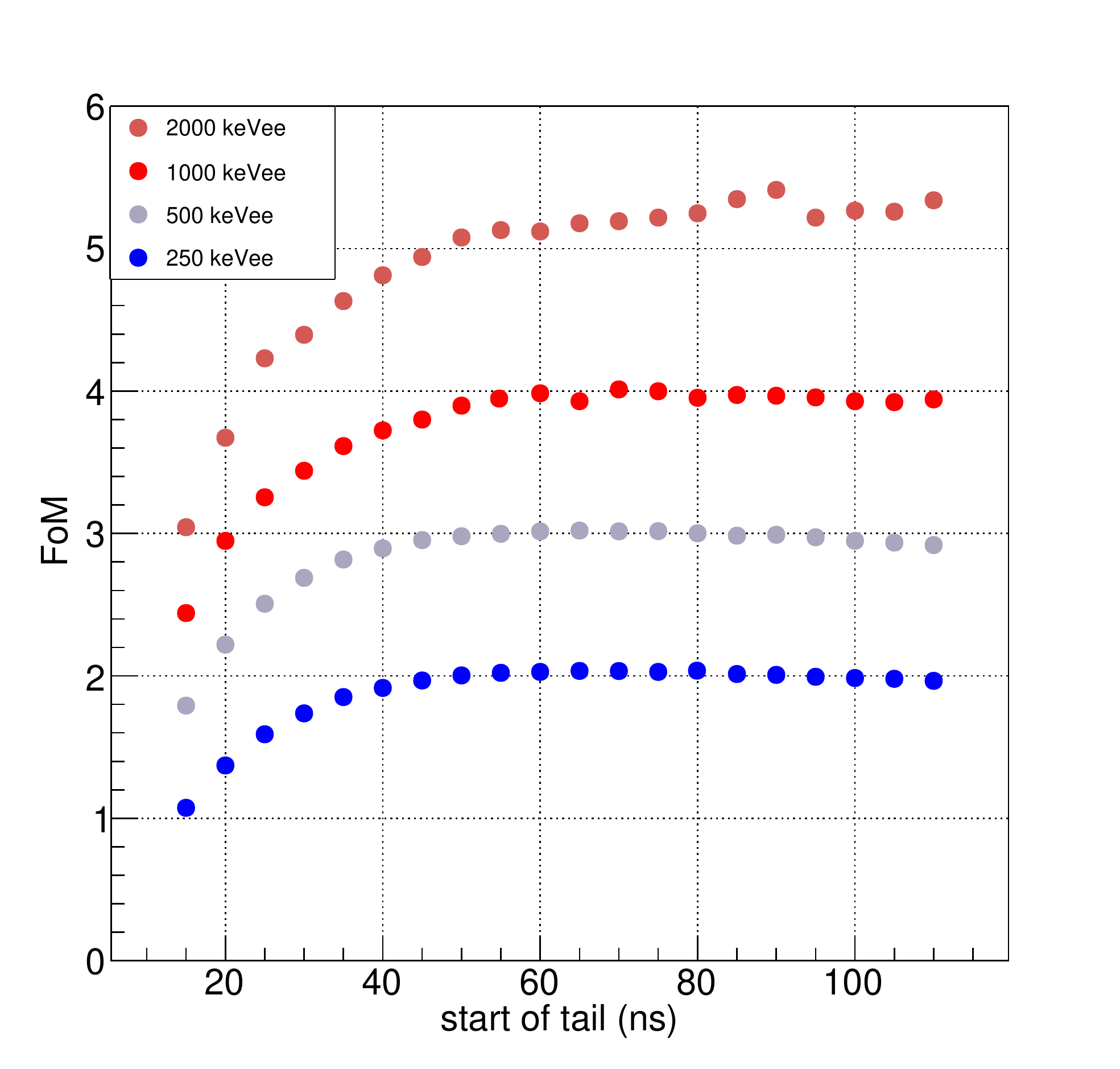}
		\caption{}
	\end{subfigure}
	\begin{subfigure}[h]{0.32\columnwidth} 
		\centering
		\includegraphics[width=\columnwidth]{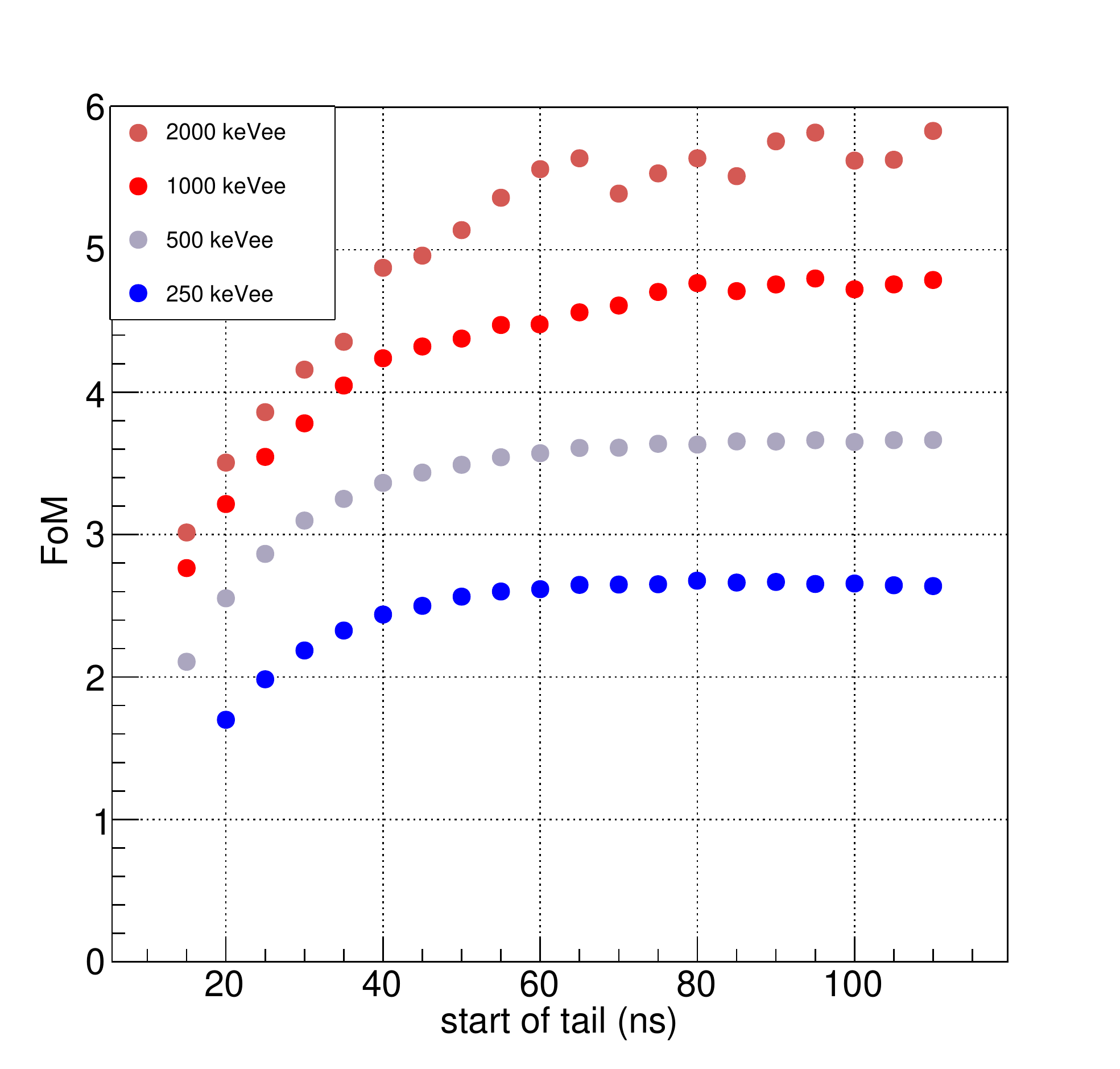}
		\caption{}
	\end{subfigure}
\caption{(a) The $FoM$ as a function of tail start time for the (a) SensL's ArrayC-60035-64P-PCB, (b) SensL's ArrayJ-60035-64P-PCB, and (c) the Hamamatsu S13361-6050 array.}
\label{fig:fig8} 
\end{figure}

\begin{figure}[!htbp]
	\begin{subfigure}[h]{0.32\columnwidth} 
		\centering
		\includegraphics[width=\columnwidth]{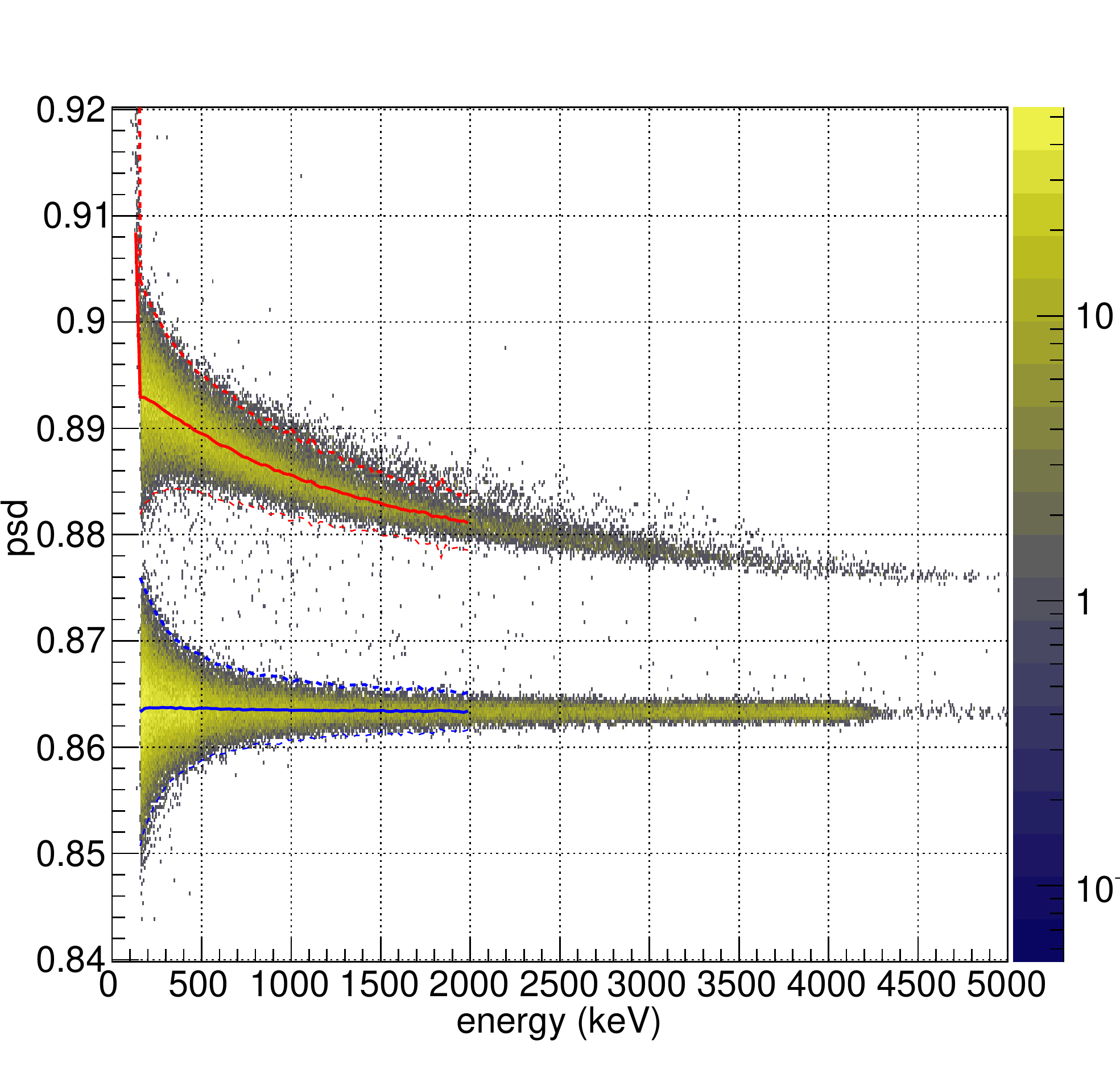}
		\caption{}
	\end{subfigure}	
	\begin{subfigure}[h]{0.32\columnwidth} 
		\centering
		\includegraphics[width=\columnwidth]{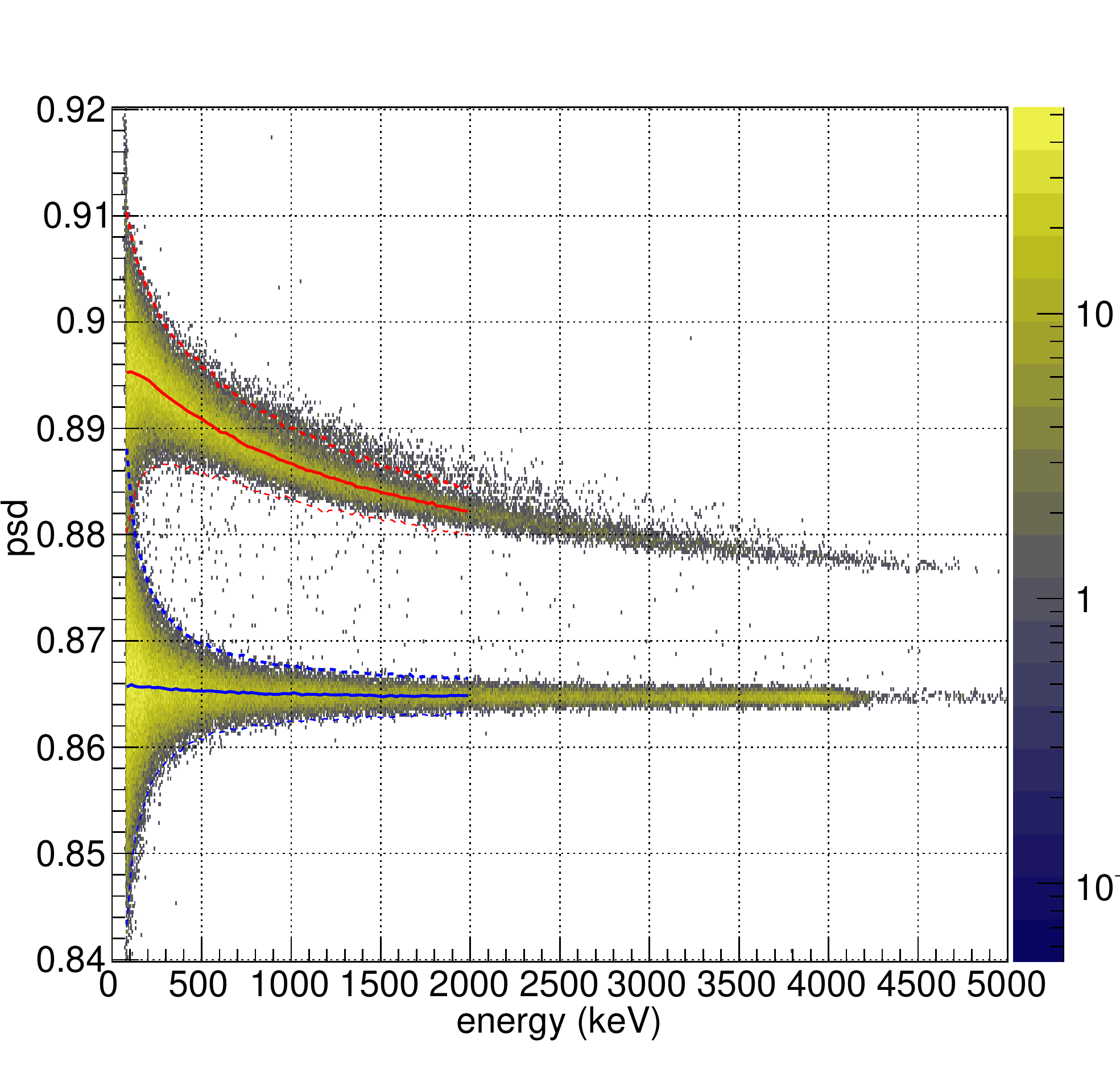}
		\caption{}
	\end{subfigure}
	\begin{subfigure}[h]{0.32\columnwidth} 
		\centering
		\includegraphics[width=\columnwidth]{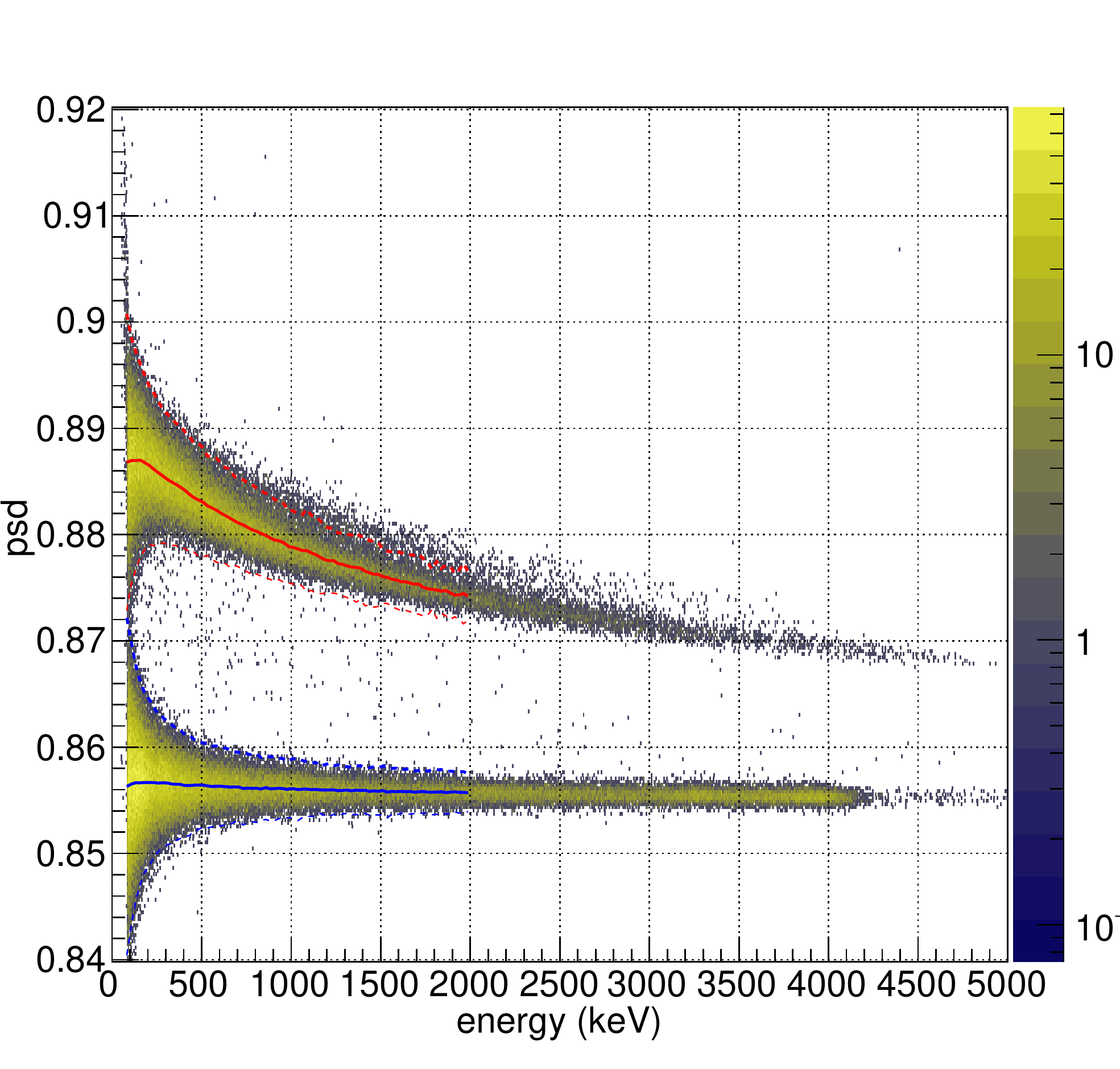}
		\caption{}
	\end{subfigure}
\caption{The optimized pulse-shape discrimination parameter as a function of energy for (a) SensL's ArrayC-60035-64P-PCB, (b) SensL's ArrayJ-60035-64P-PCB, and (c) the Hamamatsu S13361-6050 array. }
\label{fig:fig9} 
\end{figure}

\begin{figure}[!htbp]
	\centering
	\includegraphics[width=0.5\columnwidth]{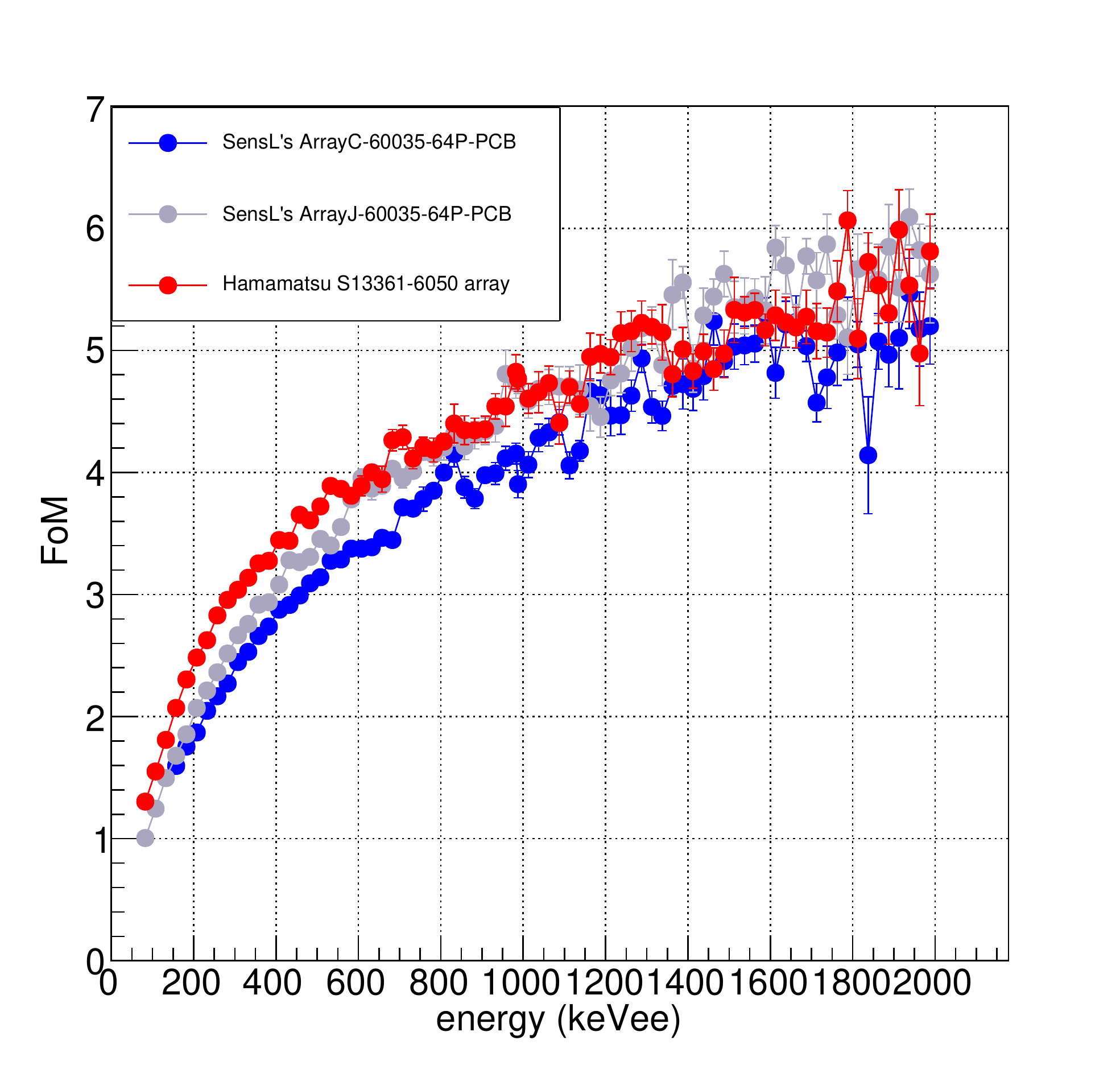}
\caption{The $FoM$ for a 60 ns tail start time as a function of energy for all SiPM arrays. The errors are the standard deviation of the results of fits to 20 bootstrapped distributions obtained by sampling with replacement.}
\label{fig:fig10} 
\end{figure}

The pulse-shape parameter used in this work is defined as the ratio of the tail pulse integral to the total pulse integral. The start time of the tail region is determined by optimizing the figure-of-merit, defined as
\begin{equation}
FoM = \frac{\mu_n-\mu_\gamma}{2.355(\sigma_n+\sigma_\gamma)},
\end{equation}
where $\mu_{n/\gamma}$ is the mean of a Gaussian fit to the neutron and gamma distributions, $\sigma_{n/\gamma}$ is the standard deviation of the Gaussian fit, and the factor of 2.355 is for conversion to a FWHM width. Figure \ref{fig:fig8} shows the $FoM$ as a function of tail start time for all three SiPM arrays. The shape of the fission neutron energy spectrum favors optimizing the $FoM$ for lower energy depositions in order to maximize overall neutron detection efficiencies. For this reason, a window of 60 ns was chosen as the tail integral start time. At this value and for each array, the 250 keVee $FoM$ curves have maximized. The $FoM$ for 2000 keVee depositions for the J-series and Hamamatsu appears to continue to increase for later tail start times: this effect is especially pronounced for the Hamamatsu array. The optimized PSD parameter as a function of energy for all PMTs is shown in Figure \ref{fig:fig9}, and values for four different energies are listed in Table \ref{tab:tab3}: the mean of the gamma and neutron populations are indicated by solid lines (red for neutrons and blue for gammas), and the upper and lower (3$\sigma$) bounds are indicated by dashed lines (red for neutrons and blue for gammas). Finally, the optimized energy-dependent $FoM$ is shown in Figure \ref{fig:fig10}. The errors are the standard deviation of the results of fits to 20 bootstrapped distributions obtained by sampling with replacement. The Hamamatsu array out-performs both the J-series and C-series array until about 500 keVee. At approximately 1500 keVee, the Hamamatsu's performance is comparable to the J-series. However, with an energy dependent tail-start time, the $FoM$ at high energies can be improved.

\begin{table}

\begin{tabular}{|c||c|c|c|}
\hline
Array					&\hspace{0.9mm}\small{80-110 keVee}\hspace{0.8mm}	&\hspace{1.7mm}\small{230-260 keVee}\hspace{1.6mm}	&\hspace{1.1mm}\small{330-360 keVee} \hspace{1.1mm} 		\\
\hline
\small{ArrayC-60035-64P-PCB}		&-										&2.05$\pm$0.01							&2.53$\pm$0.02	\\
\small{ArrayJ-60035-64P-PCB}		&1.01$\pm$0.01							&2.21$\pm$0.03							&2.76$\pm$0.02	\\	
\small{Hamamatsu S13361-6050}		&1.31$\pm$0.01							&2.63$\pm$0.02							&3.14$\pm$0.04	\\	
\hline
\end{tabular}
\begin{tabular}{|c||c|c|c|}
\hline
Array					&\small{455-485 keVee}		&\small{1035-1065 keVee}		&\small{1985-2015 keVee}			\\
\hline
\small{ArrayC-60035-64P-PCB	}	&2.99$\pm$0.03		&4.28$\pm$0.11		&5.20$\pm$0.31			\\
\small{ArrayJ-60035-64P-PCB	}	&3.27$\pm$0.04		&4.68$\pm$0.20		&5.62$\pm$0.40			\\	
\small{Hamamatsu S13361-6050}		&3.65$\pm$0.03		&4.66$\pm$0.17		&5.81$\pm$0.31			\\	
\hline
\end{tabular}
\caption{The PSD $FoM$ for different energy depositions for each array. The errors are the standard deviation of 20 fit trials in which the original distribution was sampled with replacement. The threshold for the ArrayC-60035-64P-PCB measurement was higher than 110 keVee.}
\label{tab:tab3}
\end{table}

\section{7 cm diameter measurement}
\begin{figure}[!htbp]
	\begin{subfigure}[h]{0.49\columnwidth} 
		\centering
		\includegraphics[width=\columnwidth]{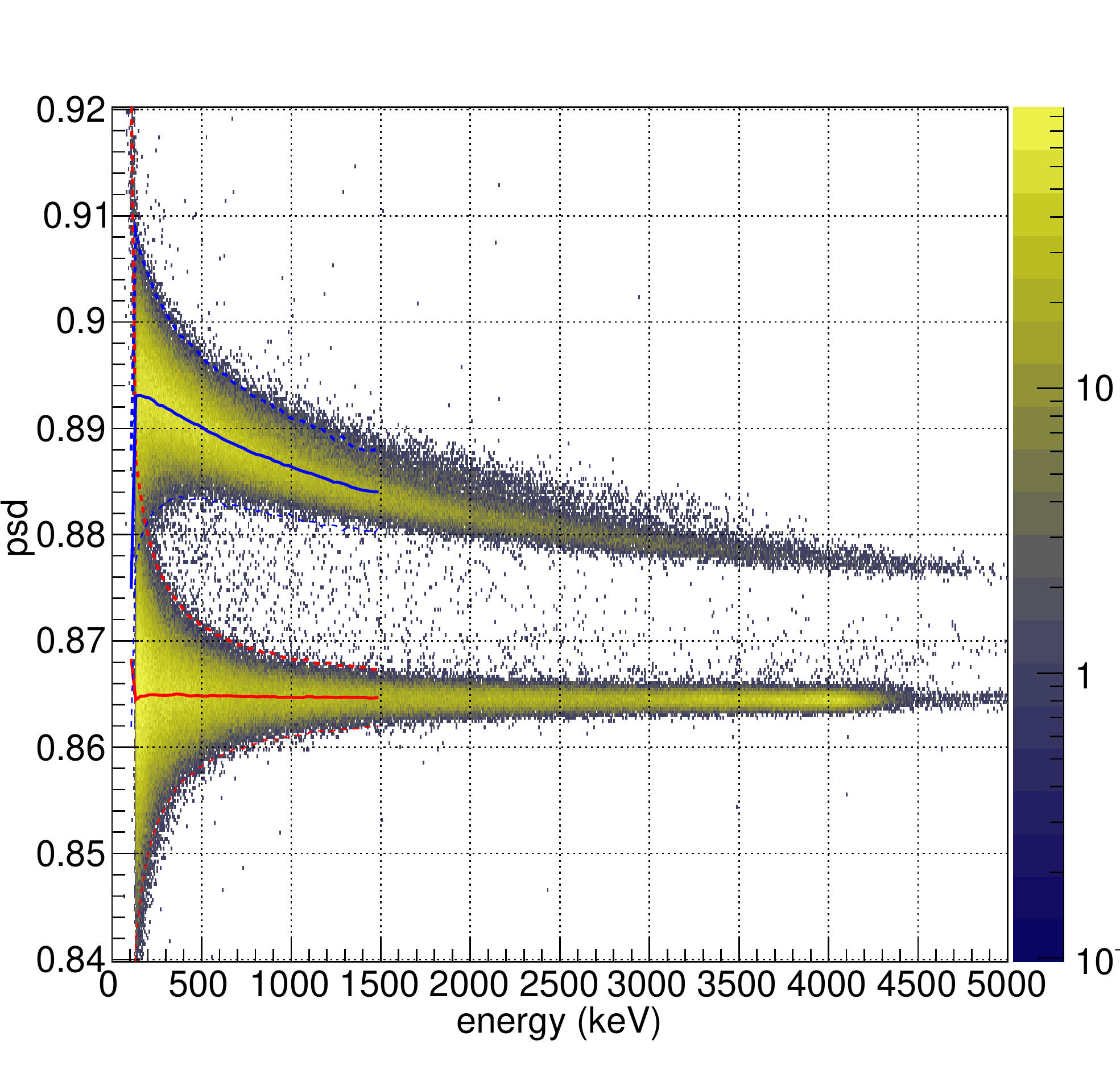}
		\caption{}
	\end{subfigure}	
	\begin{subfigure}[h]{0.49\columnwidth} 
		\centering
		\includegraphics[width=\columnwidth]{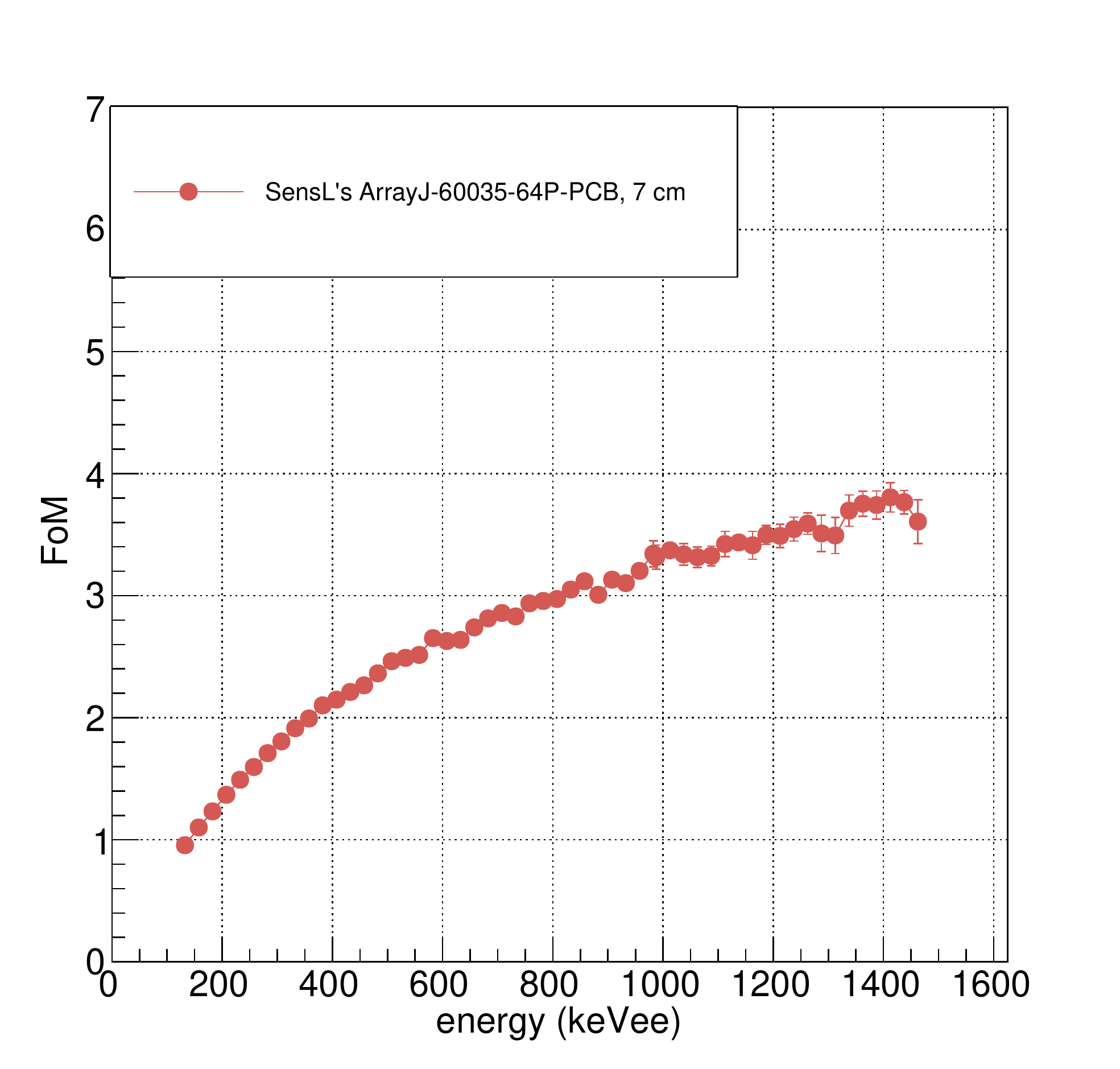}
		\caption{}
	\end{subfigure}

\caption{The optimized pulse-shape discrimination parameter (a) and $FoM$ (b) as a function of energy for SensL's ArrayJ-60035-64P-PCB coupled to the 7 cm diameter crystal. The errors are the standard deviation of 20 fit trials in which the original distribution was sampled with replacement.}
\label{fig:fig11} 
\end{figure}

A final measurement set was conducted using a 7 cm diameter cylindrical {\it{trans}}-stilbene crystal (also from Inrad Optics) coupled to the J-series array. Our results with the 5 cm diameter crystal indicated that the J-series had the best overall performance. The J-series array was coupled with optical grease to the center of the readout surface of the crystal. In this configuration, the portions of the readout surface of the crystal that were not coupled to the SiPM array were left bare, and all other surfaces were wrapped in Teflon. The results are presented in Table \ref{tab:tab4} along with results for the 5 cm diameter crystal. For the timing measurement, only one PMT was powered during the measurement. We use the average of the previous three measurements for $\sigma_{A-B}$, and use the errors for the prior J-series measurement for $\sigma_{A-T}$ and $\sigma_{B-T}$. For the PSD parameter, we found that a tail start time of 60 ns was still the optimal window for energies 500 keVee and below.  The optimized PSD parameter and $FoM$ as a function of energy are shown in Figure \ref{fig:fig11}.

\begin{table}
\centering
\begin{tabular}{|cl||c|c|c|}
\hline
 \hspace{0.4cm}Scintillator:& Array								&$\sigma_{E}$ (\%)	&$\sigma_{T}$ (ps)	&$FoM$  	\\
\hline
5 cm stilbene:	&\small{ArrayC-60035-64P-PCB}		&13.5$\pm$1.0	&471$\pm$25		&2.05$\pm$0.01				\\
			&\small{ArrayJ-60035-64P-PCB}		&13.6$\pm$1.8	&277$\pm$34		&2.21$\pm$0.03				\\
			&\small{Hamamatsu S13361-6050}	&13.7$\pm$0.9	&420$\pm$32		&2.63$\pm$0.02				\\
\hline
7 cm stilbene:	&\small{ArrayJ-60035-64P-PCB}		&16.3$\pm$0.7	&578$\pm$22		&1.49$\pm$0.01				\\

\hline
\end{tabular}
\caption{A summary of results for all arrays. The energy resolution $\sigma_E$ is fit at the Compton edge of 511 keV photons (341 keVee), the timing resolution $\sigma_T$ is for 100-400 keVee depositions, and the $FoM$ is measured at 230-260 keVee.}
\label{tab:tab4}
\end{table}

As expected due to the incomplete coverage of the crystal, all metrics are degraded to some degree compared to the measurements on the 5 cm diameter crystal. Each metric can be expected to improve by optimizing the 
SiPM coverage with a custom SiPM array or wrapping the uncoupled surface with a reflector material. 

\section{Discussion and Conclusions}
We report on the energy, timing, and pulse-shape discrimination performance of 5 cm diameter {\it trans}-stilbene crystal read out with three commercial SiPM arrays, with the individual pixel outputs passively summed into one output signal. We also report on the performance of a 7 cm diameter {\it trans}-stilbene crystal read out with the summed output of the ArrayJ-60035-64P-PCB SiPM from SensL. While large variability in PMT capabilities makes direct comparisons difficult, recent measurements of the PSD $FoM$ for a 5 cm diameter {\it trans}-stilbene crystal with a Photonis XP4512B PMT readout quotes a value of 1.8 in the 200-300 keVee energy range \cite{sosa}, 20\% higher than our 7 cm diameter measurement and 22\% lower than our 5 cm diameter measurement. Based on these results, using the passively summed output of commercially available SiPM arrays as a PMT replacement results in comparable or better performance for applications requiring good timing and PSD performance. While performance degrades due to incomplete coverage for a 7 cm crystal,  it may be acceptable for many applications. Future work will explore the energy resolution performance in inorganic scintillators, and integrate a temperature compensation circuit to account for changes in the breakdown voltage with temperature. 

\section{Acknowledgments}
The energy calibration software used in this work was written by Joshua Brown, formally of the University of California at Berkeley and currently at Sandia National Laboratories in Livermore, California.

We would like to thank UK MOD/AWE and the US DOE National Nuclear Security Administration, Office of Defense Nuclear Nonproliferation Research and Development for co-funding this work. This paper describes objective technical results and analysis. Any subjective views or opinions that might be expressed in the paper do not necessarily represent the views of the U.S. Department of Energy or the United States Government. Document Release Number SAND2019-4774 J.

Sandia National Laboratories is a multimission laboratory managed and operated by National Technology and Engineering Solutions of Sandia, LLC, a wholly owned subsidiary of Honeywell International, Inc., for the U.S. Department of Energy's National Nuclear Security Administration under contract DE-NA0003525. This paper describes objective technical results and analysis. Any subjective views or opinions that might be expressed in the paper do not necessarily represent the views of the U.S. Department of Energy or the United States Government.

\bibliographystyle{../../../LatexTools/IEEE_style/IEEEtran}
\bibliography{%
}

\appendix
\section{ArrayX-BOB6\_64S summing board readout}
\begin{figure}[!htbp]
		\centering
		\begin{subfigure}[h]{0.52\columnwidth} 
			\includegraphics[width=\columnwidth]{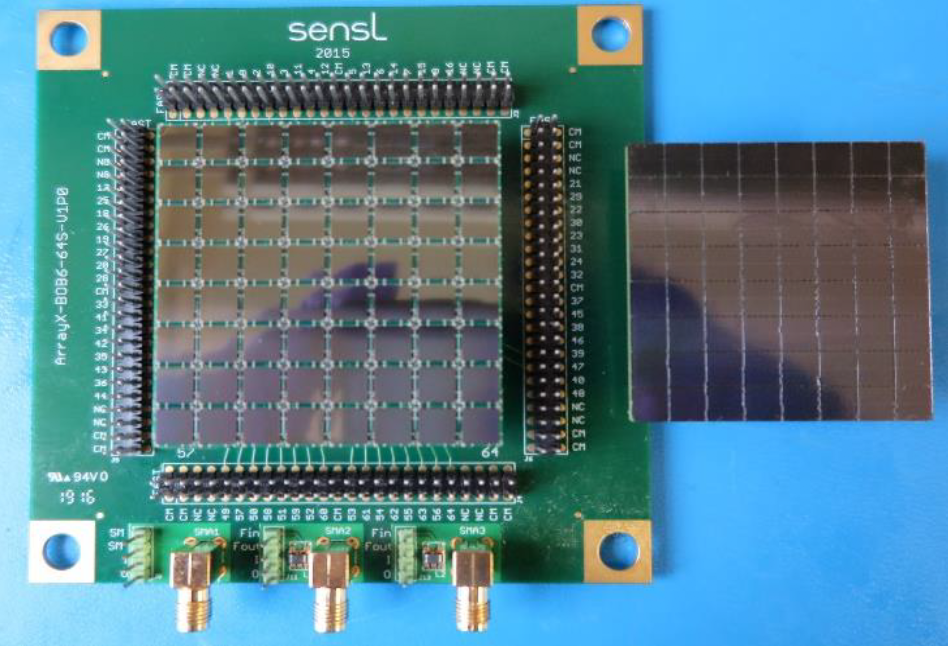}
			\caption{}
		\end{subfigure}	
		\begin{subfigure}[h]{0.42\columnwidth} 
			\includegraphics[angle=90,width=\columnwidth]{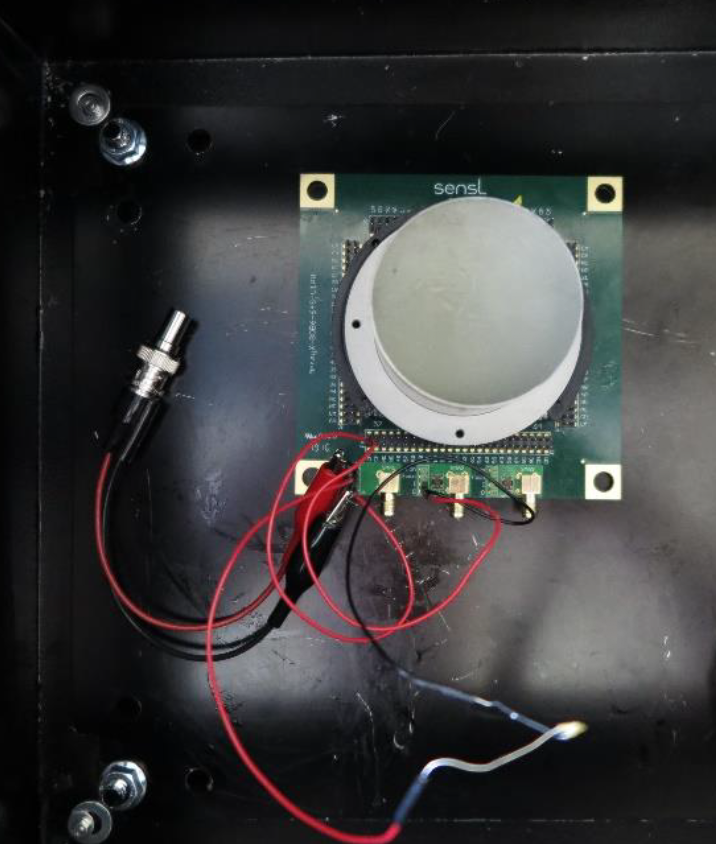}
			\caption{}
		\end{subfigure}	
	\caption{ (a) The SensL ArrayC-60035-64P-PCB SiPM mounted to the ArrayX-BOB6\_64S summing board and the SensL ArrayC-60035-64P-PCB SiPM. (b) The scintillator mounted to the ArrayC-60035-64P-PCB read out with the ArrayX-BOB6\_64S: note the associated jumper leads, resistor and capacitor required to assemble the readout circuit.}
	\label{fig:figA1} 
\end{figure}

Prior to designing and testing a passive summing board, we used SensL's  ArrayX-BOB6\_64S to determine if the pulse-shape discrimination (PSD) was comparable to a photo-multiplier tube. We compared the PSD performance from a 5 cm diameter x 5 cm long right cylinder of {\it trans}-stilbene crystal wrapped with Teflon using an ETL 9214 PMT and both the ArrayC-60035-64P-PCB and ArrayJ-60035-64P-PCB from SensL, with their outputs summed with the ArrayX-BOB6\_64S. The SensL arrays and summing board are shown in Figure \ref{fig:figA1}a. The charge pulses emitted from the anode of the ETL 9214 PMT and SiPM arrays were captured and digitized by a CAEN DT5730B digitizer. This digitizer is capable of digitizing pulses at a sampling rate of 500 MS/s with a 14-bit resolution and a selectable input dynamic range of 2 or 0.5 Vp-p. To perform PSD, a Digital Pulse-Processing Algorithm (DPP) was executed on a Field Programmable Gate Array (FPGA) based upon the digitizer. The DPP-PSD algorithm is a charge integration algorithm which integrates the digitized pulse over two time windows (the long and short gate). 

The experimental setup consisted of a $^{252}$Cf source placed 5 cm away from the scintillator detector crystal, which are subsequently mated to photo-detector of interest via EJ-550 optical grease. To reduce the gamma ray count rate to the same order as the neutrons, a 2 cm lead brick was placed between the source and detector crystal.
For the case of measurements taken with the ETL 9214 PMT, the high voltage was supplied by a CAEN N1470 high voltage supply. The bias used for each scintillator was chosen to optimize the digitization resolution. The charge pulses resulting from the Compton edge of interactions of 662 keV photons from a $^{137}$Cs calibration source interacting with the scintillator cell were observed on an oscilloscope terminated at 50 $\Omega$. The bias was adjusted the so the peak amplitude was -340 $\pm$ 20 mV on the oscilloscope. A peak amplitude of 340 mV was used as this equates to a measurable energy range of 0--3 MeVee across the 2 Vp-p dynamic range of the DT5730B digitizer. During the measurements described above, the PMT anode output was observed on an oscilloscope terminated at 1 M$\Omega$ and subsequently connected to the DT5730B digitizer for digitization and data processing. 

\begin{figure}[!htbp]
		\centering
		\begin{subfigure}[h]{0.49\columnwidth} 
			\includegraphics[width=\columnwidth]{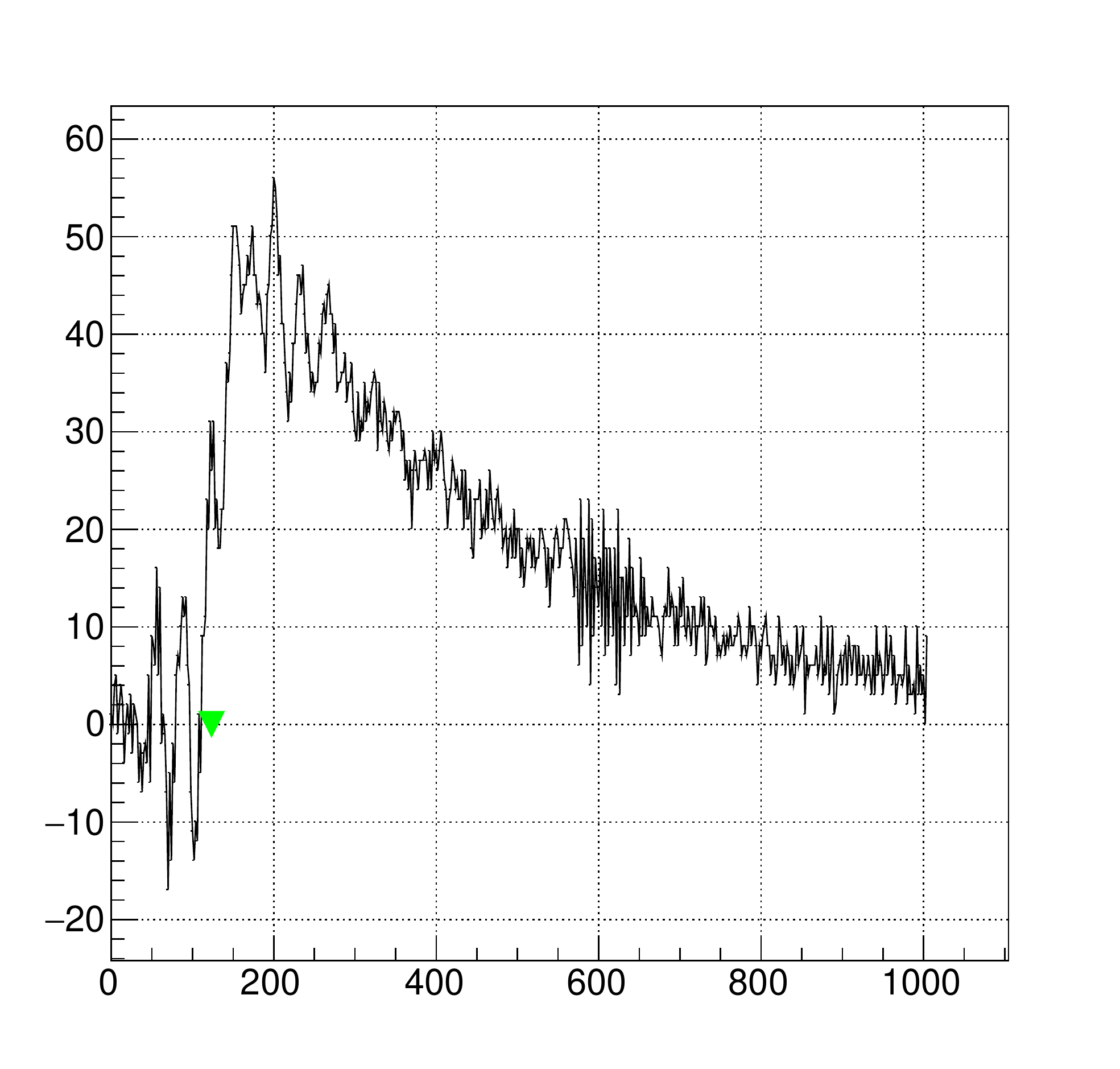}
			\caption{}
		\end{subfigure}	
		\begin{subfigure}[h]{0.49\columnwidth} 
			\includegraphics[width=\columnwidth]{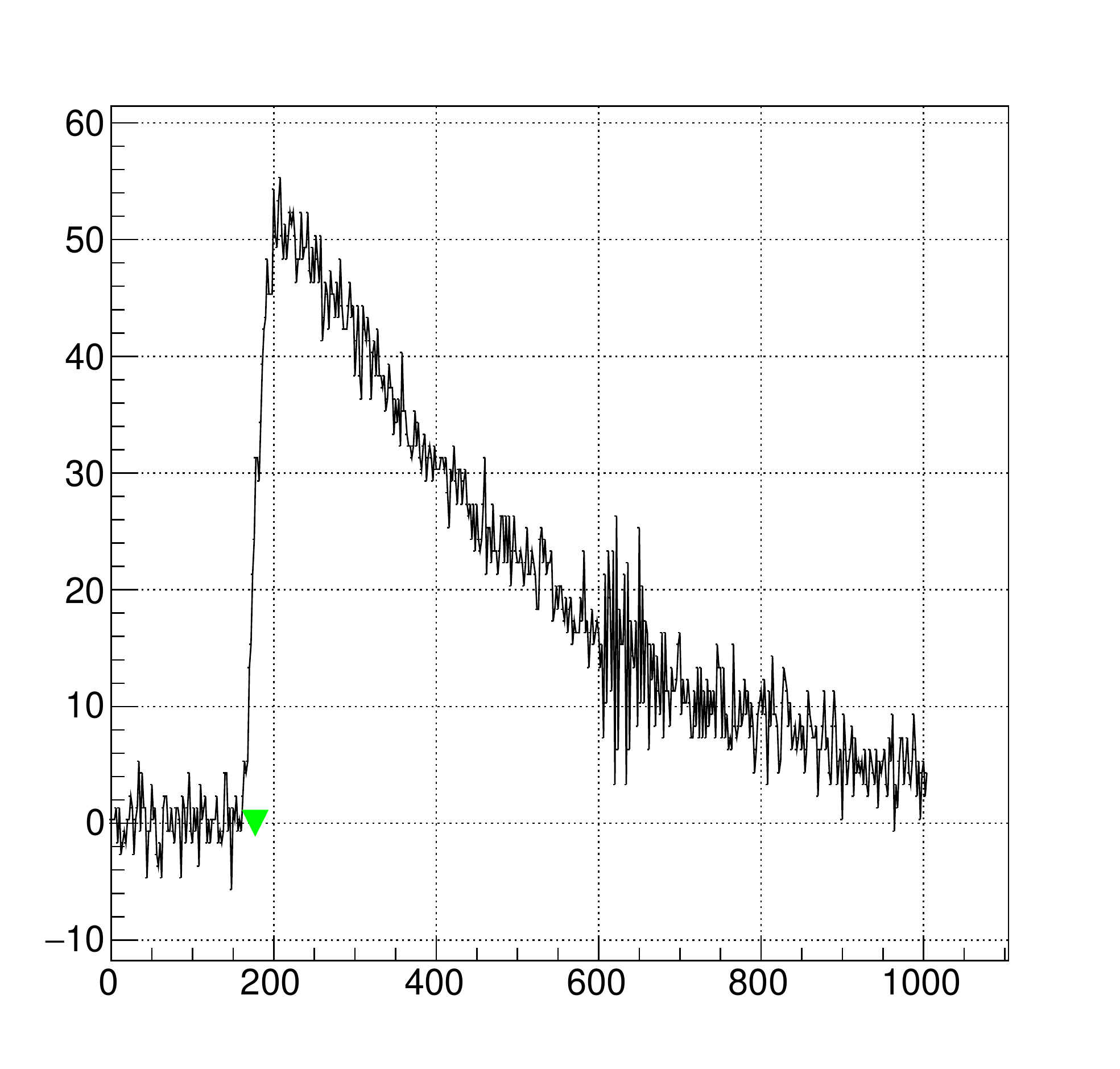}
			\caption{}
		\end{subfigure}	
	\caption{Two example traces from the ArrayX-BOB6\_64S board in an un-shielded configuration. The noise is presumably due to the lead wires required to build the readout circuit, and is greatly reduced when the assembly is placed in a shielded container. }
	\label{fig:figA2} 
\end{figure}

A photo of the experimental setup used for the SiPM array is shown in Figure \ref{fig:figA1}b. The detector is coupled to the SiPM array, which was mounted the to the summing board and biased at -29.65 V.  The sum of the SiPM signals were read through a 10 nF capacitor. To connect the 50 $\Omega$ resistor, 10 nF capacitor and the other jumper pins required for readout and bias, the 5 inch long jumper leads supplied by SensL were used. A large amount of noise can be observed due to these cables (see Figure \ref{fig:figA2}), which will influence the observed detector performance. The sum signal of the SiPM array was observed on an oscilloscope terminated at 1 M$\Omega$ and subsequently connected to the DT5730B digitizer utilizing the 0.5 Vp-p dynamic range for digitization and data processing.

The PSD parameters for the DPP-PSD algorithm will influence the measured performance of the scintillators. We recorded 350,000 pulse waveforms to optimize the PSD parameters in software. To ensure the whole scintillation pulse shape for each event was recorded, the pulse record length was 1024 ns and 2048 ns for PMT and SiPM pulses, respectively. For each set of pulses, the digitizer was configured with an 88 ns pre-trigger. The start time of the short and long gates is 40 ns prior to the trigger point. The 48 ns time period before the gate start time was used to calculate the baseline. A MATLAB code was written to optimize the PSD gate by maximizing the figure-of-merit ($FoM$) above 450 keVee for a range of short and long gate windows. 

\begin{figure}[!htbp]
			\includegraphics[width=\columnwidth]{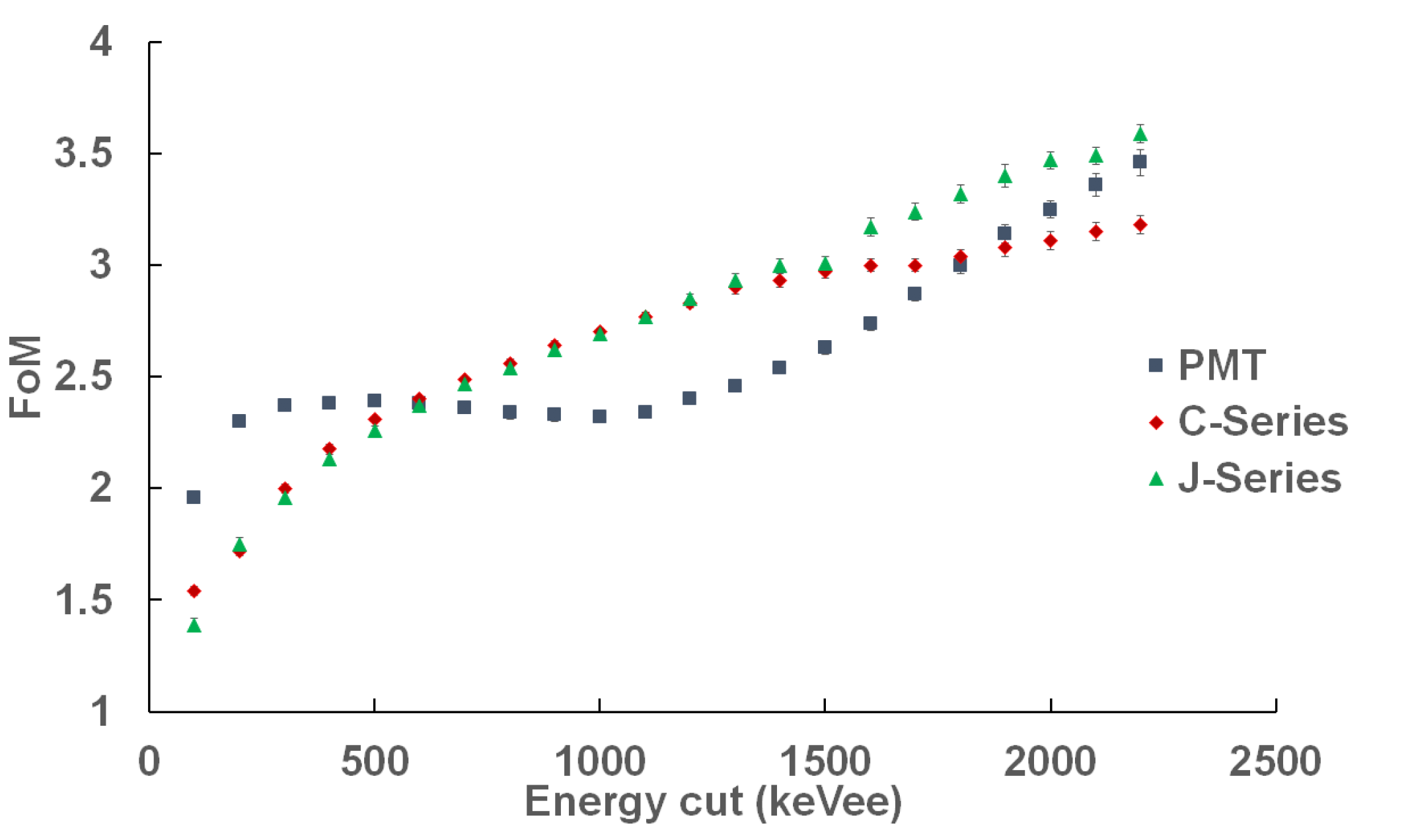}
			\caption{The PSD $FoM$ as a function of energy for a 5 cm diameter stilbene crystal scintillator coupled to an ETL 9214 PMT, a SensL ArrayC-60035-64P-PCB SiPM, and a SensL ArrayC-60035-64P-PCB SiPM. The SensL arrays are summed by SensL's ArrayX-BOB6\_64S summing board. }
	\label{fig:figA3} 
\end{figure}

Once the optimum charge integration gates have been found, 1,000,000 events resulting from the experimental setup described were recorded. Figure \ref{fig:figA3} show the $FoM$ for the ETL 9214 PMT, and the SensL ArrayC-60035-64P-PCB SiPM and the SensL ArrayC-60035-64P-PCB SiPM summed with SensL's ArrayX-BOB6\_64S summing board. Based on this preliminary data, which shows that the summed SiPM readout has comparable PSD performance to a PMT readout, we proceeded with designing our own passive summing board. 

\end{document}